\documentclass[lettersize,journal]{IEEEtran}
\usepackage{amsmath,amsfonts}
\usepackage{algorithmic}
\usepackage{algorithm}
\usepackage{array}
\usepackage[caption=false,font=normalsize,labelfont=sf,textfont=sf]{subfig}
\usepackage{textcomp}
\usepackage{stfloats}
\usepackage{url}
\usepackage{verbatim}
\usepackage{graphicx}
\usepackage{cite}
\usepackage{tabularx}
\usepackage{slashbox}
\newcommand{\ax}{x} % absolute position along the trajectory
\newcommand{\at}{t} % absolute time

\newcommand{\w}{\mathrm{e}} %exit from the motorway towards the workshop

\newcommand{\ptl}{c_1(i,s)}

\newcommand{\di}{s} % distance
\newcommand{\td}{T} % time duration 

\begin{document}

\title{Multi-criteria Decision-making of Intelligent Vehicles under Fault Condition Enhancing Public-private Partnership}

\author{Xin Tao, Mladen \v{C}i\v{c}i\'{c}, Jonas Mårtensson,~\IEEEmembership{Member,~IEEE}
        % <-this % stops a space
\thanks{X. Tao is with the Integrated Transport Research Lab, KTH Royal Institute of Technology, 10044, Stockholm, Sweden (e-mail: taoxin@kth.se).}
\thanks{M. \v{C}i\v{c}i\'{c} is with Univ. Grenoble Alpes, CNRS, Inria, Grenoble INP, \mbox{GIPSA-lab}, 38000 Grenoble, France (e-mail: cicic.mladen@gipsa-lab.fr).}
\thanks{J. Mårtensson is with the Division of Decision and Control Systems and the Integrated Transport Research Lab, KTH Royal Institute of Technology, 10044 Stockholm, Sweden (e-mail: jonas1@kth.se).}
}

% The paper headers
\markboth{Journal of \LaTeX\ Class Files,~Vol.~14, No.~8, August~2021}%
{Shell \MakeLowercase{\textit{et al.}}: A Sample Article Using IEEEtran.cls for IEEE Journals}

%\IEEEpubid{0000--0000/00\$00.00~\copyright~2021 IEEE}
% Remember, if you use this you must call \IEEEpubidadjcol in the second
% column for its text to clear the IEEEpubid mark.

\maketitle

\begin{abstract}
With the development of vehicular technologies on automation, electrification, and digitalization, vehicles are becoming more intelligent while being exposed to more complex, uncertain, and frequently occurring faults. In this paper, we look into the maintenance planning of an operating vehicle under fault condition and formulate it as a multi-criteria decision-making problem. The maintenance decisions are generated by route searching in road networks and evaluated based on risk assessment considering the uncertainty of vehicle breakdowns. Particularly, we consider two criteria, namely the risk of public time loss and the risk of mission delay, representing the concerns of the public sector and the private sector, respectively. A public time loss model is developed to evaluate the traffic congestion caused by a vehicle breakdown and the corresponding towing process. The Pareto optimal set of non-dominated decisions is derived by evaluating the risk of the decisions. We demonstrate the relevance of the problem and the effectiveness of the proposed method by numerical experiments derived from real-world scenarios. The experiments show that neglecting the risk of vehicle breakdown on public roads can cause a high risk of public time loss in dense traffic flow.
With the proposed method, alternate decisions can be derived to reduce the risks of public time loss significantly with a low increase in the risk of mission delay. This study aims at catalyzing public-private partnership through collaborative decision-making between the private sector and the public sector, thus archiving a more sustainable transportation system in the future.
\end{abstract}


\section{Introduction} \label{sec1}
% maintenance decision-making is important
Vehicles are highly complex systems that are prone to errors, faults, and system failures. The rapid development of vehicular technologies on automation, electrification, and digitalization is making the vehicles more intelligent and sustainable, but also bringing about more faults, such as anomalous sensors \cite{realpe2015sensor}, faulty actuators \cite{ren2021traditional} and battery issues \cite{lu2013review}. These faults could be complex, uncertain, and highly recurring, thus threatening the reliability, safety, and efficiency performance of the vehicles. Various techniques have been developed to obtain more accurate fault information, such as fault detection and diagnosis \cite{chen2019review} \cite{shafi2018vehicle} and fault prognosis \cite{li2015prediction}. To enable vehicles under fault condition to continue driving, considerable studies have been done on fault tolerant design, which is still under active exploration \cite{spooner1997fault} \cite{du2021fault}. Despite these efforts, unless the faulty component is repaired or replaced, the vehicle is still exposed to the risk of unexpected failures, leading to a vehicle breakdown and downtime. 

For long-distance freight deliveries with tight schedules, the vehicles undergo long operational hours and little planned downtime during the mission. If a faulty component deteriorates and causes a system failure and a vehicle breakdown, both the road traffic and the profit of the logistics company are at risk. Under fault condition, the time or distance a vehicle can drive until the breakdown is uncertain due to many factors, such as uncertain root cause of the fault \cite{TAOFAULT} and uncertain degrading process \cite{de2020aging}. In this situation, the decision-making of the vehicle regarding how to replan the delivery mission and the maintenance is critical while challenging.

% human can do it, but not good enough
According to a recent field study in \cite{9613093}, for a human-driven truck under fault condition, today's practice relies on the driver to make a decision or communicate with remote personnel regarding whether to seek maintenance before finishing the delivery mission. This human decision-making process is limited for several reasons: 
\begin{itemize}
    \item The expertise of the driver on vehicular faults is limited, especially for faults arising from new technologies.
    \item The communication between the driver and remote personnel is limited by the availability of the remote personnel and the quality of their dialogue. 
    \item For a highly automated vehicle, the driver has limited responsibility for its operation \cite{TaxonomyAD}. For a fully autonomous vehicle, the driver is not even onboard.
    \item The advancement of sensing and data storage technologies makes more information and data available, which are difficult for humans to interpret and utilize.
\end{itemize}

Therefore, we need to automate this decision-making process and make it more reliable, efficient, and intelligent. In \cite{tao2022short}, we conducted preliminary research on developing a short-term maintenance planning model for a truck under fault condition, which is limited mainly to two aspects. Firstly, the decision alternatives are largely simplified and manually defined without considering the vehicle routing problem in a road network. More importantly, we took the minimization of the economic risk of the private sector as the objective of the maintenance planning. Although it is a common practice today, a vehicle breakdown may also bring high risk to society.

As a typical cause of traffic bottlenecks, a vehicle breakdown may expose public transport to a high risk of traffic congestion. It was reported in \cite{ma2017prioritizing} that approximately 60$\%$ of traffic congestion delay is triggered by non-recurrent incidents. Statistics also showed that vehicle breakdowns accounted for about 30$\%$ of all types of road incidents in New South Wales in Australia \cite{chand2020analysis} and 64$\%$ of all incidents on a motorway in the United Kingdom \cite{wang2005vehicle}. In \cite{chand2020analysis}, a steep rise in vehicle breakdown frequency on motorways was disclosed in a case study in New South Wales. Despite these facts, there has not been a study in the traffic management domain on how an individual vehicle breakdown and the towing process may affect traffic flow and cause traffic congestion. 

% we see an gap here
This mutual lack of consideration between the public sector and the private sector provides an opportunity to enhance public-private partnership. By considering the impact on public transport, we can reconstruct the maintenance planning process of an operating vehicle toward a more sustainable solution. The potential for collaborative decision-making between traffic management authorities and autonomous vehicles (AVs) has attracted increasing attention. As proposed in \cite{taha2018route}, such a collaboration may lead to safer and faster resolution of congestion in instances of traffic management during events or emergencies. In \cite{trid_2020}, 'Will AVs increase or decrease total traffic flow and congestion?' is proposed as a critical question for the future autonomous vehicle landscape. It also mentions that public-private partnership is a way to go for introducing AVs into public transport systems. Although this potential has been identified, the ways to implement it are yet to be developed. 

% we are doing something here
In this paper, we consider the situation in which a vehicle during a delivery mission detects a fault and is likely to break down before the delivery. We formulate the maintenance planning of the vehicle as a multi-criteria decision-making (MCDM) problem considering both the concerns of the public sector and the private sector. Specifically, based on probability theory and risk assessment, we evaluate the risk of the decisions with two criteria, namely the risk of public time loss and the risk of mission delay, and derive the Pareto front as the optimal decision set. The main contributions of the paper are as follows.
\begin{itemize}
    \item {We formulate an MCDM problem for the maintenance planning of an operating vehicle under fault condition considering the concerns of both the public sector and the private sector.} 
    \item {We build a public time loss model to evaluate the traffic congestion caused by a vehicle breakdown considering the dynamic traffic flow and the towing process of the vehicle.} 
    \item {We build a risk evaluation model of uncertain vehicle breakdowns that enables independent development of a public time loss model and a mission delay model.} 
    \item {We build a decision-alternative generator by route searching in road networks, which automates the identification of decision alternatives.} 
    \item {We establish numerical experiments for a typical long-distance freight delivery scenario where the decision-making of the vehicle is executed in different situations.}
\end{itemize}

The remainder of the paper is organized as follows. In Section \ref{sec2}, a brief literature review is presented. Section \ref{sec3} formulates the MCDM problem with details and describes the overall flow of the proposed method. Section \ref{sec4} presents the method of solving the MCDM problem in detail with mathematical models. Section \ref{sec5} provides numerical experiments, results, and discussions. Section \ref{sec6} discusses the application of the proposed method and its limitations. Section \ref{sec7} draws conclusions and provides some suggestions for future work.

\section{Relevant work} \label{sec2}
In the context of this paper, the decision of the vehicle includes the execution sequence of maintenance and delivery and the relevant routes. In the MCDM problem, the impact of vehicle breakdown on traffic congestion is a critical criterion to evaluate, which is a challenge this paper tackles. Overall, there are mainly four domains or perspectives involved, including maintenance planning, vehicle routing problem, traffic congestion management, and MCDM. We provide a brief review of their relevant studies below. 

% maintenance planning 
Maintenance planning is crucial for the availability of machines and is widely considered for a wide range of applications in transportation, including automobile \cite{shafi2018vehicle}, railway \cite{wu2018optimizing}, ships \cite{ellefsen2019comprehensive} and so on. The challenges and opportunities of vehicle maintenance are gaining increasing attention in research communities, especially with the development of AV technologies. A recent literature review of maintenance approaches revealed the need for detailed quantitative analyses to properly choose the maintenance strategy \cite{nowakowski2018evolution}. Analyses in \cite{vskerlivc2020analysis} showed that scientific research that deals with heavy vehicle maintenance, especially modern truck maintenance guidelines are lacking. In \cite{tavares2016vehicles}, the necessity to study the use of emerging technologies in vehicle maintenance was identified, especially data processing and communications technologies. Therefore, it is important to tackle the emerging challenges of vehicle maintenance by taking advantage of new technologies.

In recent years, benefiting from the development of sensing and prognosis technologies, maintenance strategies are transitioning from reactive to proactive and are becoming more adaptive and intelligent \cite{ellefsen2019comprehensive}. As a typical and fast-developing proactive strategy, predictive maintenance takes the advantage of real-time condition monitoring and prognosis of equipment and provides more accurate maintenance plans \cite{voronov2020machine} \cite{prytz2014machine}. Predictive maintenance has received wide attention and is becoming the common practice for critical and costly components in the vehicle industry \cite{shafi2018vehicle}. As a key enabler of predictive maintenance, prognosis technology can predict the development of the fault and estimate the remaining useful life of the equipment \cite{shafi2018vehicle} \cite{ahmadzadeh2014remaining}. It is becoming increasingly advanced with the support of various data-driven methods, especially machine learning methods \cite{ellefsen2019comprehensive}. In this paper, it is also a prerequisite that the vehicle being considered is equipped with fault prognosis technologies. 

Research on the maintenance planning of vehicles has been conducted from various perspectives. In \cite{bouvard2011condition}, an optimization of grouping maintenance operations for a commercial heavy vehicle was addressed to reduce the global maintenance cost of the system. In \cite{8790142}, a vehicle fleet maintenance scheduling optimization problem was solved by a multi-objective evolutionary algorithm. \cite{biteus2017planning} used constraint programming for flexible maintenance planning and route optimization for fleet utilization optimization for heavy trucks. Despite these efforts, these studies mostly considered known health states of vehicles and focused on the global maintenance performance of fleets. In other words, unexpected faults and the risk of vehicle breakdown during a transport mission are rarely considered. Accordingly, the mobility and the vehicle routing problem of the vehicle at risk of breakdowns in a road network are not considered, which limits the applicability of predictive maintenance for long-freight delivery vehicles with high availability requirements. 

% logistics engineering 
In logistics engineering, vehicle breakdown is regarded as one of the four main types of disruptions and is considered in disrupted vehicle routing problems (DVRP). The review in \cite{eglese2018disruption} showed that DVRP for road freight transport has been a widely studied area for years. Various types of breakdown disruption are considered in DVRP, including determined, dynamic, and random breakdowns. In \cite{ahmadi2013location}, a location-routing problem in a supply-chain network was considered where the vehicles involved in the distribution system are disrupted randomly. In \cite{abu2021development}, a multilayered agent-based heuristic system was developed for vehicle routing problem under random vehicle breakdown. In these studies, regardless of the breakdown type, the DVRP was considered only after the breakdown happens, while an operating vehicle at risk of breakdown was not covered. Furthermore, in DVRP, a spare vehicle is usually used to take over the delivery mission of the broken down vehicle, while vehicle maintenance before delivery is not considered \cite{eglese2018disruption}. In \cite{dhahri2013vehicle}, although maintenance was considered in a DVRP, it focused on the rescheduling of the supply plan in case of planned preventive maintenance instead of maintenance itself. However, at risk of breakdown, getting maintenance first and delivering afterward is a strategy that is used in practice \cite{9613093}. In the end, most of the targets of planning and decision-making in DVRP are related to the profit of the private sector, while the impact of vehicle disruption on the public has not been considered \cite{eglese2018disruption}. 

% traffic management
In the traffic congestion management domain, vehicle breakdown and slow vehicles are typical road incidents that cause traffic congestion. There are extensive and profound studies on how such traffic bottlenecks are formed \cite{li2020fifty} and on how traffic congestion can be estimated \cite{akhtar2021review} and relieved \cite{nguyen2020traffic}. Among them, the time-space diagram is at the core of many traffic flow theory innovations. It is a comprehensive diagram containing all the information required to estimate the microscopic and macroscopic characteristics of the traffic stream \cite{khajeh2019traffic}. However, according to \cite{chand2021examining}, many previous studies have not differentiated the road incident types, and studies that focus on the specific type of vehicle breakdown are rare. Furthermore, according to our literature review, there has not been study on how vehicle breakdown (including the towing process) impact traffic flow or how to evaluate this impact. 

% MCDM
In this paper, both the concerns of the public and private are considered for criteria that might conflict with each other, which is a typical MCDM problem. There is extensive research on MCDM with various solutions, such as Particle Swarm Optimization \cite{wu2018optimizing}, evolutionary algorithm \cite{8790142}, and local search \cite{bouziyane2020multiobjective}. However, these methods have limited applicability in solving the problem in this paper, particularly for two reasons:
\begin{itemize}
    \item With a fault detected on a vehicle, the distance the vehicle can drive is probabilistic and this probability is continuous. Thus the search methods that require determined or discrete vehicle states are less applicable. 
    \item The impact of vehicle breakdown on traffic flow is a dynamically evolving process, the evaluation of which is complex. As a result, it is difficult to approximate this impact with a single loss function and combine it with other loss functions. 
\end{itemize}

Therefore, in this paper, we use risk assessment to evaluate each criterion of each solution, thus preserving the continuous property of vehicle breakdown probability while keeping the evaluation of different criteria separate. Risk-based decision-making is widely adopted in research, especially in handling uncertainty \cite{aven2013uncertainty}. Since it is a new tryout in this paper to consider both the public and private sectors, there is no consensus or practice to refer to regarding the trade-off between the two. Therefore, in this paper, the set of optimal decisions, instead of a unique optimal decision, is provided considering multiple criteria. For this purpose, obtaining the Pareto front is a feasible solution, which refers to deriving the set of all non-dominated solutions and is a well-adopted method in multi-objective optimization \cite{bouziyane2020multiobjective} \cite{biswas2019multiobjective}. The solutions in this set are non-dominated by each other but are superior to the rest of the solutions in the decision space.

\section{Problem formulation and method overview}\label{sec3} 
\subsection{Problem formulation} \label{sec3.1}
% motivate the scope of the problem 
In a typical vehicle routing problem for freight deliveries, the vehicle is routed to the customer starting from the warehouse. Assume that the vehicle is detected with a fault on the road and may break down before finishing the delivery. In this situation, a maintenance mission is presented, which can be executed before, simultaneously with, or after the delivery mission. For conciseness, we refer to the vehicle with the delivery mission as the ego vehicle to distinguish it from the other vehicles that may be involved, such as the tow truck. We refer to the time and the location where the fault is detected as the alarm time and the alarm location, respectively. 

In practice, depending on the specific situation and available resources, there are various ways to continue the delivery and execute the maintenance. The delivery mission can be executed by a spare vehicle or by the ego vehicle after maintenance. The maintenance can be executed onsite by an assistant vehicle or in a workshop. The ego vehicle can request to be towed to a workshop or drive to a workshop itself. As discussed in the literature review, usually, a spare vehicle or a tow truck is used only after a vehicle breakdown. The expenses of using spare vehicles, assistant vehicles, or tow trucks are all economic losses to the private sector instead of the public one. With these considerations, we formulate the problem with two simplifications: 
\begin{itemize}
    \item {The ego vehicle executes both the delivery and the maintenance. A spare vehicle taking over the delivery mission is not considered. }
    \item {To be maintained, the ego vehicle drives to the workshop itself. Towing service is only used after a vehicle breakdown.}
\end{itemize}

With the first simplification, we reduce the complexity of introducing a spare vehicle as an additional agent that causes private loss. In this way, we focus on multi-criteria decision-making considering both the public sector and the private sector. With the second simplification, we reserve the uncertainty of vehicle breakdown, which is another focus of this paper. Since these simplifications only involves some specific and detailed aspects, they do not change the key properties of the problem itself. 

\begin{figure}[htb]
        \centering
        \includegraphics[width=0.45\textwidth]{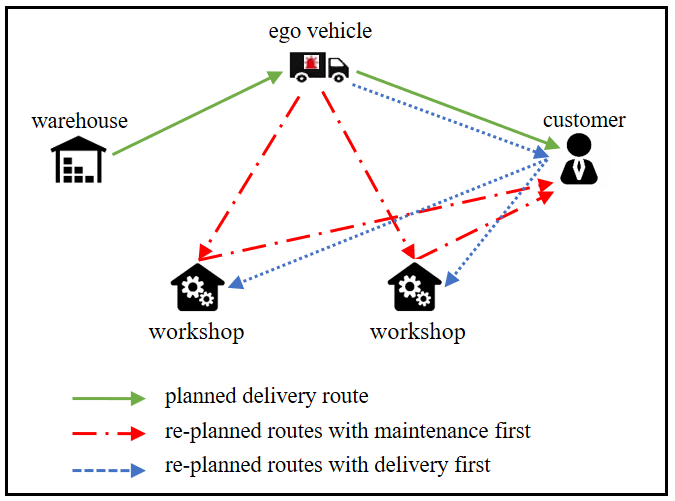}
        \caption{Problem visualization.}
        \label{fig:1}
\end{figure}

The problem is briefly visualized in Fig. \ref{fig:1}. The ego vehicle has a planned route from a warehouse to a customer, which takes the shortest time to deliver and is determined before the mission is executed. When a fault is detected, it can continue the delivery mission and be maintained afterwards, shown as the blue routes. It can also drive to one of the available workshops for maintenance first and deliver afterward, shown as the red routes. Depending on the road network, each arrow in the figure may represent one route or a group of routes with a specific origin and destination pair. 

To decide to deliver first or maintain first, we consider the concerns of both the public sector and the private sector. For the public sector, the main concern is the traffic congestion a vehicle breakdown and the corresponding towing process may cause. We represent this concern with the risk of public time loss as the one decision criterion. Note that there might be other relevant criteria, we choose the public time loss since it involves both the spatial and temporal properties of traffic congestion. For the private sector, the main concern is if the delivery mission can be finished on time. Therefore, we represent this concern with the risk of mission delay in time as the second criterion. 

To be evaluated, a feasible decision needs to specify: a) the execution sequence of maintenance and delivery; b) which workshop for maintenance; and c) all the relevant routes. Therefore, a decision can be represented by a sequence of two routes. The information of a) and b) is contained in the ending nodes of each of the two routes. To keep the focus of the paper on dealing with the uncertainty of vehicle breakdowns and the multi-criteria perspective of the decision-making problem, we further assume that: 
\begin{itemize}
    \item {If the ego vehicle is maintained in a workshop before delivery, it drives to the customer with the route that takes the shortest time. In this situation, the ego vehicle is in a determined healthy state.}
    \item {If the ego vehicle breaks down before being maintained, a worst-case strategy is applied where \textit{a workshop schedules a tow truck and tows the ego vehicle back for maintenance so that the ego vehicle leaves the public road in the shortest time}. In this situation, the ego vehicle is in a determined broken-down state.}
    \item {If the ego vehicle delivers first and reaches the customer without breakdown, it drives to a workshop for maintenance so that the risk of public time loss is minimal. In this situation, the risk of mission delay is no longer a relevant criterion.} 
\end{itemize}

To this end, we formulate the maintenance planning of an operating vehicle in the risk of breakdown as an MCDM problem. The decision space is a set of route sequences, with each sequence containing two ordered routes. The criteria of decision-making are the risk of public time loss and the risk of mission delay.
    
\subsection{Method overview}\label{sec3.2}

\begin{figure}[htb]
        \centering
        \includegraphics[width=0.45\textwidth]{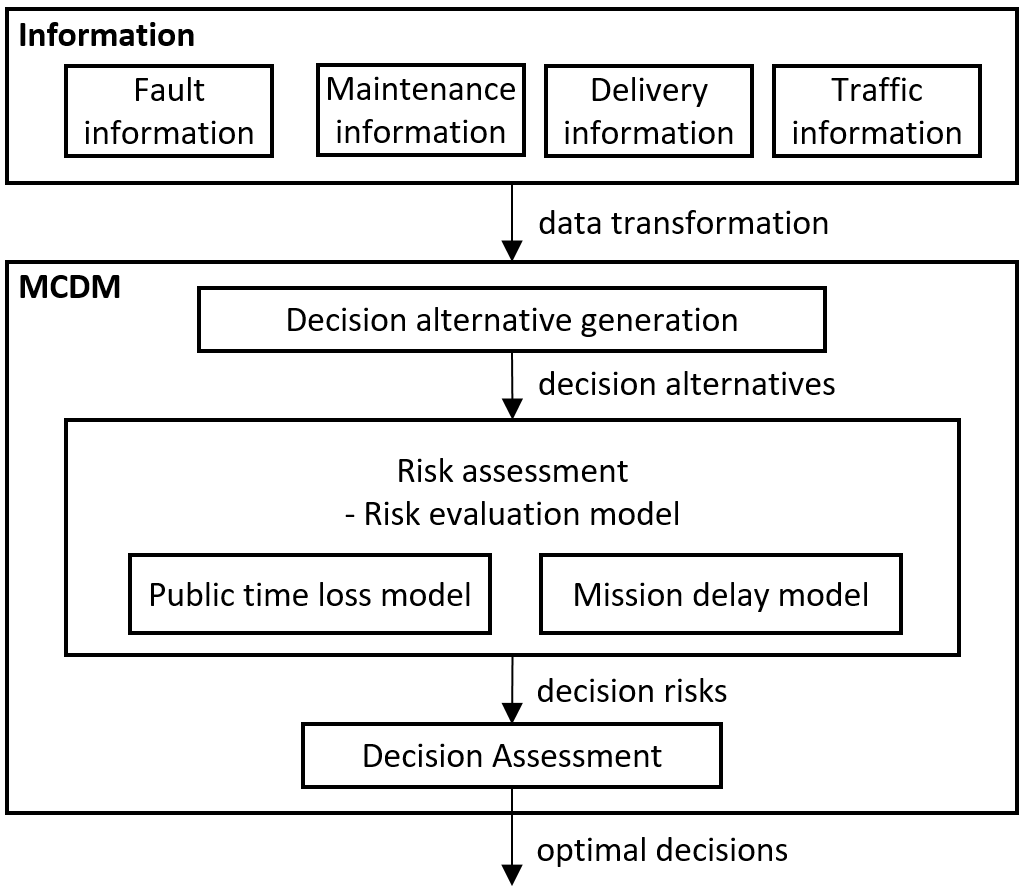}
        \caption{Method overview.}\label{fig:2}
\end{figure}

We propose a risk-based decision-making method for the MCDM problem, with an overview shown in Fig. \ref{fig:2}.

Various types of information are required as inputs to the decision-making model, which can be divided into four main categories:
\begin{itemize}
\item \textbf{Fault information}, including the alarm time, the alarm location, and the remaining useful life (RUL) of the ego vehicle. The RUL is supposed to be provided by the fault prognosis module and can be either the time or the distance the vehicle is able to drive until breakdown depending on the specific prediction model. Due to the uncertainty of the degradation process of the faulty component, the RUL is also uncertain, which was explained with details in \cite{tao2022short}. In this paper, we assume the RUL of the vehicle is represented with the remaining drivable distance and follows a probability distribution, which is predicted by other models based on sensor data or historical data.
\item \textbf{Maintenance information}, including the locations of available workshops, the maintenance time of the ego vehicle, and the tow truck service information. The maintenance time of the ego vehicle depends on many factors, such as the specific fault types, the expertise of the workshops, and the availability of spare parts. In this paper, this information is an input to the decision-making model and is assumed to be provided by the workshops. 
\item \textbf{Traffic information}, including the estimated real-time traffic flow on public roads, the road traffic capacities, and the speed limits of roads. The estimation of traffic flow is well supported by model-based methods or data-driven methods, such as convolutional neural network \cite{khajeh2019traffic}.
\item \textbf{Delivery information}, including the mission deadline.  
\end{itemize}

Note that there can be other relevant types of information in each of the information categories and those that are not directly used in the paper are not listed. Moreover, these information categories are not exclusively independent. For example, the maintenance time of the ego vehicle may depend on the fault type of the vehicle. 

The different types of information are interpreted, transformed into data and fed into the decision-making model. In the MCDM model, the decision alternatives are generated. Then the risk of the decisions is quantitatively evaluated by a risk evaluation model, including the risk of public time loss and risk of mission delay. The risk evaluation model consists of two low-level models, the public time loss model, and the mission delay model. In the end, the performances of the decision alternatives are assessed based on the risk values and the optimal decisions is obtained. 

Note that as an MCDM problem, there may be multiple optimal decisions. How to choose a single optimal decision among multiple ones is not covered in this paper, but we provide a discussion in Section \ref{sec6.1}. It is also worthy to mention that in this paper, risk evaluation emphasizes the quantitative measuring of risk, while risk assessment refers to the process of identifying and analyzing potential hazardous events, evaluating the consequences of these events, and evaluating the risk of taking different measures. 

\section{Method}\label{sec4}

In this section, the method is implemented following the steps shown in Fig. \ref{fig:2}, including decision-alternative generation, risk assessment, and decision assessment. 

\subsection{decision-alternative generation} \label{sec4.1}

%%% === As I mentioned earlier in an e-mail, try being more consistent with the notation. The less symbols are "recycled" the better. It can be good to compile a list of the main notation, even if you don't use it in the paper, just as a reference to yourself. For example, if you use $l$ for "loss", don't use it in superscripts. Using it as $\textrm{l}$ in superscript helps, but it's better to avoid "notation clashes" altogether.

%%% === If you want to use $l$ as the length of the road (although $d$ might work better for "distance"), you should redefine the notation for loss. Perhaps $\lambda$ for loss might work?

%%% === "Free flow" speed is the widely accepted term for what you call the "normal" speed in the traffic community. I think this should be rephrased throughout the paper (see my edits). I suppose the notation should also change if the term is changed, since ^n would no longer make sense with "free flow" instead of "normal".

%%% === I've changed the notation from ^n (representing "normal" speed / time) to ^{\mathrm{ff}} (representing "under free flow conditions").

We model the road network as a directed graph ${G}=({N}, {E}, {T^{\mathrm{ff}}})$, where ${N}$ is the node set, $ {E}$ is the link set, and $T^{\mathrm{ff}}$ is the set of link travel times at free flow speed. The node set ${N}$ consists of three types of nodes with ${N}= {N}^{\mathrm{w}} \cup  {N}^{\mathrm{tr}} \cup {N}^{\mathrm{cu}} $, where ${N}^{\mathrm{w}}$ is the set of workshop nodes, ${N}^{\mathrm{tr}}$ is the set of traffic nodes representing intersections, and ${N}^{\mathrm{cu}}$ is the set of the customer node. Two nodes $n_1$ and $n_2$ in $N$ are adjacent if there is a link $(n_1, n_2) \in E$ or $(n_2, n_1) \in E$. The free flow travel time on a link is computed as the travel time without any congestion, driving at free flow speed.

The alarm location of the ego vehicle is represented by two attributes: the link it travels on $(n^{\textrm{ua}}, n^{\textrm{da}})$ and the distance $d^\textrm{ua}$ to node $n^{\textrm{ua}}$. Here $n^{\textrm{ua}}$ and $n^{\textrm{da}}$ refer to the upstream node and the downstream node of the alarm location respectively. By adopting existing route searching algorithms, such as depth-first search \cite{tarjan1972depth}, the set of routes from the downstream node $n^{\textrm{da}}$ to workshop nodes in $ {N}^{\mathrm{w}}$ can be obtained, denoted as $R^{\textrm{w}}$. If a route in $R^{\textrm{w}}$ contains the node in ${N}^{\mathrm{cu}}$, it is removed from $R^{\textrm{w}}$ since it passes the customer and can deliver first. Similarly, the set of routes to the customer node in ${N}^{\mathrm{cu}}$ can be obtained, denoted as $R^{\textrm{cu}}$. These two route sets make up the whole route set of the first mission $R^{\textrm{1}}$, i.e. $R^{\textrm{1}}=R^{\textrm{w}} \cup R^{\textrm{cu}}$. 

A decision $r$ is a sequence of two routes, denoted as $r = \{r^{\textrm{1}}, r^{\textrm{2}}\}$, where $r^{\textrm{1}}$ is the route of the first mission and $r^{\textrm{2}}$ is the route of the second mission. Under the assumptions in the problem formulation,  each $r^{\textrm{1}}$ corresponds to a unique  $r^{\textrm{2}}$ . Therefore, the total number of decisions is the number of routes in $R^{\textrm{1}}$, denoted as $I$. The $i_{th}$ decision $r_i= \{r_i^{\textrm{1}}, r_i^{\textrm{2}}\}$, with  $r_i^{\textrm{1}}=R^{\textrm{1}}(i), i \in [1,I]$.

For a decision with $r_i^{\textrm{1}} \in R^{\textrm{w}}$, the ego vehicle drives to the workshop first. After the maintenance in the workshop, it drives to the customer with the route that takes the shortest time. By adopting existing shortest path search algorithms, such as Dijkstra's algorithm and A* algorithm \cite{rachmawati2020analysis}, the route of the delivery mission $r_i^{\textrm{2}}$ can be obtained in road network $G$.

For a decision with $r_i^{\textrm{1}} \in R^{\textrm{cu}}$, the ego vehicle drives to the customer first. After the delivery, it drives to a workshop so that the risk of public time loss is minimal. The route to the workshop $r_i^{\textrm{2}}$ involves risk evaluation of public time loss, which is presented in the next subsection. 

For any decision, before being maintained, the ego vehicle is likely to break down. If it happens, based on the worst-case strategy, a workshop will send a tow truck to the breakdown site and tow the ego vehicle back for maintenance, so that the ego vehicle leaves the public road as soon as possible. According to traffic regulations, a tow truck towing another vehicle drives at a reduced speed, thus taking more time on each link. We build another road network ${G'}=( {N}, {E}, {T^{\textrm{tt}}})$, where ${N}$ and $ {E}$ are the same as in $G$, and $T^{\textrm{tt}}$ is the set of link travel times of towing.

Given that the ego vehicle takes decision $r_i$ and breaks down after driving a distance $\di$ from the alarm location, the node of the workshop that sends the tow truck and conducts maintenance can be obtained by solving an optimization problem 
\begin{equation}
    n^* = \mathop {\textrm{arg} \min}\limits_{n\in {N}^{\mathrm{w}}} \hspace{1mm}\td^{\mathrm{wb}}_{n}(i,\di|G) + \td^{\mathrm{bw}}_{n}(i,\di|G')
\end{equation}
where $\td^{\mathrm{wb}}_{n}(i,\di|G)$ refers to the shortest time the tow truck drives from the workshop $n$ to the breakdown site in $G$, and $\td^{\mathrm{bw}}_{n}(i,\di|G')$ refers to the shortest time the tow truck tows the ego vehicle to workshop $n$ in $G'$. Given the value of $i, \di, n$ and the configurations of $G$ and $G'$, both of them can be obtained by shortest path searching algorithms. 

To this end, given the alarm time and the alarm location, the way to generate all the relevant routes that constitute the decision alternatives are presented. 

\subsection{Risk assessment}\label{sec4.2}
In this subsection, we perform a risk assessment of the decisions, including building a risk evaluation model, a public time loss model, and a mission delay model. 

\subsubsection{Risk evaluation model}
As discussed in Section \ref{sec3.2}, we assume that the remaining drivable distance $S$, representing the RUL of the ego vehicle, follows a probability distribution with a probability density function $f_S(\di)$ and a corresponding cumulative distribution function $F_S(\di)= \int_{- \infty}^\di f_S(\di) \mathop{d\di}$.

For decision $r_i$ with $r_i^{\textrm{1}}\in R^{\textrm{w}}$, the ego vehicle drives to the workshop for maintenance first. It might break down before or upon reaching the workshop. Denote the two criteria, namely the public time loss and the mission delay as $j=1$, and $j=2$, respectively. The risk of $j$ giving $r_i$ can be modeled as 
\begin{equation}
     RS_j(i)=  \int_0^{D_i^{\mathrm{1}}}f_S(\di)c_j(i,\di)\mathop{d\di} +
    \int_{D_i^{\mathrm{1}}}^{+\infty} f_S(\di)c_j(i,\di)\mathop{d\di} 
\end{equation}
\noindent where ${D_i^{\mathrm{1}}}$ refers to the length of route $r_i^{\textrm{1}}$. $c_j(i,\di)$ refers to the loss function of $j$ given $i$ and $\di$. 

The second integral in (2) corresponds to the situation that the ego vehicle does not break down until reaching
the workshop. In this situation, $c_j(i,\di)$ is dependent from $x$. As a result, 
\begin{equation}
     \int_{D_i^{\mathrm{1}}}^{+\infty} f_S(\di)c_j(i,\di)\mathop{d\di} =
     c_j(i,\di)\int_{D_i^{\mathrm{1}}}^{+\infty} f_S(\di)\mathop{d\di}
\end{equation}
where 
\begin{equation}
    \int_{D_i^{\mathrm{1}}}^{+\infty} f_S(\di)\mathop{d\di} =1-F_S(D_i^{\mathrm{1}})
\end{equation}

For decision $r_i$ with $r_i^{\textrm{1}} \in  R^{\textrm{cu}} $, the ego vehicle delivers first. After the delivery, it drives to a workshop for maintenance. It might break down before reaching the customer, before or upon reaching the workshop. The risk of $j$ giving $r_i$ can be modeled as
\begin{equation}
\begin{split}
     RS_j(i)= & \int_0^{D_i^{\mathrm{1}}}f_S(\di)c_j(i,\di)\mathop{d\di} +  \int_{D_i^{\mathrm{1}}}^{D_i^{\mathrm{1}}+D_i^{\mathrm{2}}}f_S(\di)c_j(i,\di)\mathop{d\di} \\
     &+ \int_{D_i^{\mathrm{1}}+D_i^{\mathrm{2}}}^{+\infty} f_S(\di)c_j(i,\di)\mathop{d\di}
\end{split}
\end{equation}
where $D_i^{\mathrm{2}}$ is the length of route $r_i^{\textrm{2}}$.  

In (5), the value of $D_i^{\mathrm{2}}$ depends on $r_i^{\textrm{2}}$, which is yet to be determined. As assumed in Section \ref{sec3.1}, if the ego vehicle delivers first and reaches the customer without breakdown, it drives to a workshop for maintenance so that the risk of public time loss is minimal. By route searching in $G$, we can obtain the set of routes from the customer to the workshops, denoted as $R^{\mathrm{cw}}$. The route $r_i^{\textrm{2}}$ to go to the workshop can be obtained by solving an optimization problem 
\begin{equation}
\begin{split}
    r_i^{\textrm{2}} = & \mathop {\textrm{arg} \min}\limits_{r \in R^{\mathrm{cw}}} \int_{D_i^{\mathrm{1}}}^{D_i^{\mathrm{1}}+D(r)}f_S(\di)c_{1}(i,\di)\mathop{d\di} + \\ & \int_{{D_i^{\mathrm{1}}+D(r)}}^{+\infty} f_S(\di)c_{1}(i,\di)\mathop{d\di} 
\end{split}
\end{equation}
where $D(r)$ is the length of route $r\in R^{\mathrm{cw}}$.

The third integral in (5) corresponds to the situation that the ego vehicle does not break down until reaching the  workshop. In this situation, $c_j(i,\di)$ is independent from $\di$. As a result, 
\begin{equation}
     \int_{D_i^{\mathrm{1}}+D_i^{\mathrm{2}}}^{+\infty} f_S(\di)c_j(i,\di)\mathop{d\di} =
     c_j(i,\di)\int_{D_i^{\mathrm{1}}+D_i^{\mathrm{2}}}^{+\infty} f_S(\di)\mathop{d\di}
\end{equation}
where 
\begin{equation}
    \int_{D_i^{\mathrm{1}}+D_i^{\mathrm{2}}}^{+\infty} f_S(\di)\mathop{d\di} =1-F_S(D_i^{\mathrm{1}}+D_i^{\mathrm{2}})
\end{equation}

In (2) and (5), $c_j(i,\di)$ is a low-level loss function set containing the loss functions of different combinations of $j, i$ and $\di$. The model of the public time loss function $c_1(i,\di)$ and the model of the mission delay function $c_2(i,\di)$ are developed separately in the following subsections. Note that these models can be developed in different ways. The presented ones focus on integrating traffic information with maintenance information and providing feasible solutions.

\subsubsection{Public time loss model}  \label{sec4.2.1}

%%% === New material

If the ego vehicle breaks down on the road, it creates a zone where the traffic flow is interrupted, creating a stationary bottleneck.
This effect is present whether or not there is a shoulder lane available, albeit in that case it is less pronounced, since the vehicle does not block a traffic lane to the full.
After the tow truck arrives, it tows at a low speed, thus acting as a moving bottleneck. 
Loss functions $c_{1}(i,\di)$ involve the influence of vehicle breakdown and traffic bottleneck, potentially causing public time loss, as is modeled below.

Given decision $r_i$ and the driving distance until breakdown $\di$, the breakdown time $\at^{\mathrm{b}}(i,\di)$, the towing start time $\at^{\mathrm{t}}(i,\di)$, and the time of exiting motorways $\at^{\textrm{w}}(i,\di)$ can be computed as
\begin{align}  
    \at^{\mathrm{b}}(i,\di) &= \at^{\mathrm{a}} + \td^{\mathrm{ab}}(i,\di),\\
    \at^{\mathrm{t}}(i,\di) &= \at^{\mathrm{b}}(i,\di) + \td^{\mathrm{wb}}_{n^*}(i,\di|G),\\
    \at^{\mathrm{e}}(i,\di) &= \at^{\mathrm{t}}(i,\di) + \td^{\mathrm{be}}_{n^*}(i,\di|G^\prime),
\end{align}
where $\at^{\mathrm{a}}$ is the alarm time. The driving time from the alarm location to the breakdown location, $\td^{\mathrm{ab}}(i,\di)$, can be derived from the alarm location configured with ($n^{\textrm{ua}}$, $n^{\textrm{da}}$) and $d^{\textrm{ua}}$, the decision $r_i$, driving distance $\di$ and the free flow speed
$V^{\mathrm{ff}}_e$ on link $e \in E$. $\td^{\mathrm{be}}_{n^*}(i,\di|G^\prime)$ is the time the tow truck drives from the breakdown site to the exit of the motorway. Time duration $\td^{\mathrm{wb}}_{n^*}(i,\di|G)$ and $\td^{\mathrm{be}}_{n^*}(i,\di|G^\prime)$ are derived by solving the optimization problem in (1).

In order to estimate the public time loss, we model the evolution of the overall traffic along the trajectory of the ego vehicle.
All model parameters are determined based on $r_i$ and $\di$, which we omit for conciseness.  
We assume the entirety of the traffic affected by the ego vehicle breakdown follows the same route as it, that the net flows of vehicles entering and leaving the roads along that trajectory is negligible, and that the road capacity is constant along the ego vehicle trajectory.

\begin{figure}[t!]
        \centering
        \includegraphics[width=0.5\textwidth]{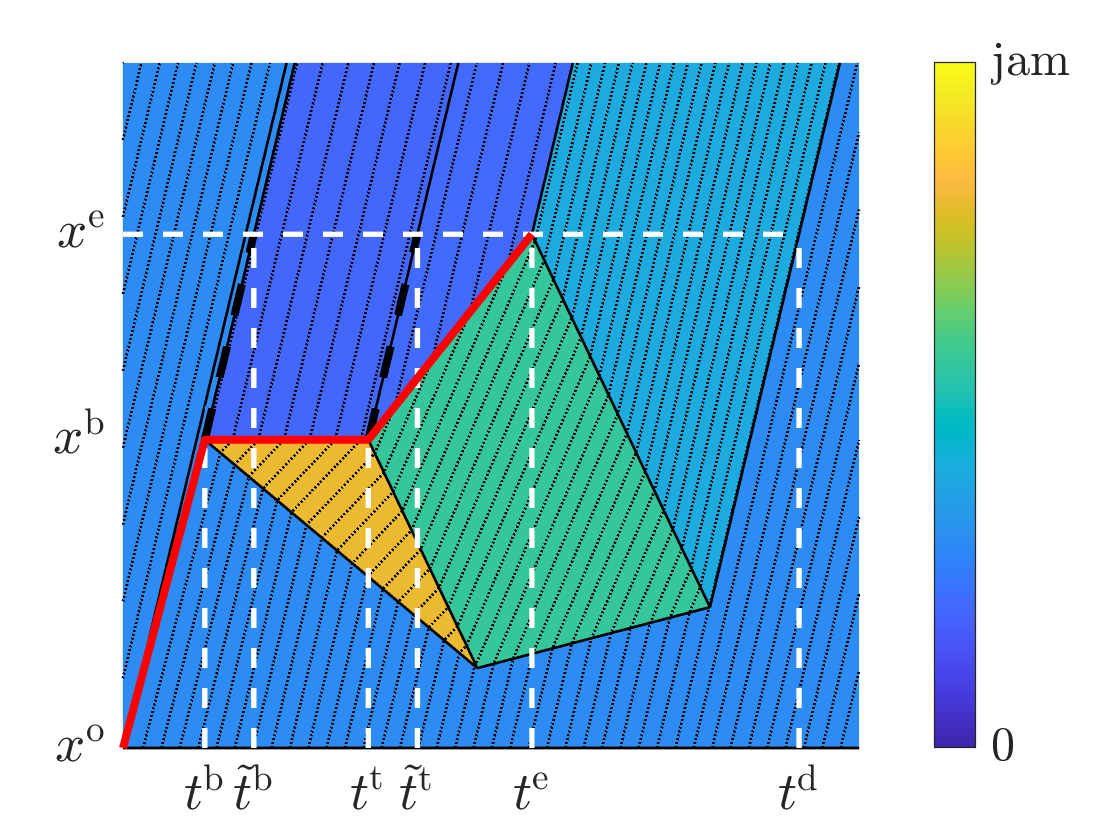}
        \caption{Time-space diagram of the traffic situation under ego vehicle breakdown disruption. Traffic density is shown colour-coded, with warmer colours representing denser traffic.}\label{fig:breaktraffic} 
\end{figure}

If the traffic flow is close to road capacity, a vehicle breakdown is likely to cause traffic congestion, regardless of if there is a shoulder lane available or not. %, since the broken down vehicle will act as a stationary bottleneck, blocking one lane.
Even after a tow truck arrives and starts towing the ego vehicle, they will still restrict the traffic flow, acting as a moving bottleneck due to their slow towing speed.
The situation in question is demonstrated in Fig.~\ref{fig:breaktraffic}, where the trajectory of the ego vehicle is shown in red.
Trajectories of a selection of other vehicles are shown in dotted black.
Once the ego vehicle breaks down, acting as a stationary bottleneck while it is waiting to be towed, other vehicles that enter the congestion that it caused (shown by the orange triangle) are forced to slow down.
The congestion that is left upstream of the ego vehicle being towed is less severe, but still causes other vehicles to move slower than in free flow.

The total public time loss due to ego vehicle breakdown can be defined as
\begin{equation} 
\label{eq:timeloss_def}
    \ptl \!= \!\!\int_{-\infty}^{\infty}\int_{-\infty}^{t} \!\!\left( \!q_{\mathrm{o}}\!\!\left(\tau - \frac{\ax^{\w}-\ax^{\mathrm{o}}}{V^{\mathrm{ff}}}\right) - q_{\w}(\tau)\! \right) \mathrm{d}\tau \mathrm{d}t,
\end{equation}
where $q_{\mathrm{o}}(t)\le Q^{\max}$ is the traffic flow through some point $\ax^{\mathrm{o}}$ upstream of the the caused congestion, $Q^{\max}$ is the capacity of the road, and $q_{\w}(t)$ is the traffic flow through point $\ax^{\w}$ where the towed ego vehicle leaves the motorway to head for the workshop.

Assuming the vehicles move at constant speed $V^{\mathrm{ff}}$ in free flow, if no bottlenecks are present on the road, the traffic flow reaching $\ax^{\w}$ is equal to the delayed traffic flow through $\ax^{\mathrm{o}}$,
\begin{equation}
    q_{\w}(\at) = q_{\mathrm{o}}\left(\at - \frac{\ax^{\w}-\ax^{\mathrm{o}}}{V^{\mathrm{ff}}}\right),
\end{equation}
and the public time loss is $\ptl=0$.
%which is not greater than the capacity of the road $Q^{\max}$.
Since the effects of ego vehicle breakdown only reflect on $q_{\w}(\at)$ after some delay, depending on the distance from the ego vehicle to $\ax^{\w}$, we use the notation $\tilde{\at}$ to specifically refer to the time when that happens.
In particular, the traffic flow overtaking the ego vehicle (both while it is stationary and while it is towed) reaches $\ax^{\w}$ at time
%We are interested in modelling the influence of the queue on the downstream flow at $\ax=\ax^{\textrm{w}}$, so the queue models the congestion upstream of the bottleneck at time $\tilde{\at}$,
\begin{equation}
\tilde{\at} = \at + \frac{\ax^{\w} - \ax_{\mathrm{ego}}(t)}{V^{\mathrm{ff}}},
\end{equation}
where $\ax_{\mathrm{ego}}(t)$ is the trajectory of the ego vehicle.
For example, the traffic overtaking the ego vehicle at times $\at^{\mathrm{b}}$ (when it breaks down) and $\at^{\mathrm{t}}$ (when it starts being towed) reach $\ax^{\w}$ at at times $\tilde{\at}^{\mathrm{b}}$ and $\tilde{\at}^{\mathrm{t}}$, respectively, as shown in Fig.~\ref{fig:breaktraffic}, following the dashed black lines.

Starting with time $\tilde{\at}^{\mathrm{b}}$, the traffic flow reaching $\ax^{\w}$ can be modelled as the output of a queueing server $q^{\mathrm{out}}(\tilde{\at})$,
\begin{equation}
\label{eq:qout}
    q_{\w}(\tilde{\at}) = q^{\mathrm{out}}(\tilde{\at}),
\end{equation}
using a variant of the queueing model presented in \cite{cicic2021coordinating}.
The evolution of the queue length ${\eta}(\tilde{\at})$, representing the aggregate influence of the congestion, is given by
\begin{equation}
\dot{\eta}(\tilde{\at}) = q^{\mathrm{in}}(\tilde{\at}) - q^{\mathrm{out}}(\tilde{\at}), \quad  \tilde{\at} \ge \tilde{\at}_{\mathrm{b}},
\end{equation}
with ${\eta}(\tilde{\at}) = 0$ for $\tilde{\at} \le \tilde{\at}^{\mathrm{b}}$.
%Due to this modelling formality, and also due to the implicit assumption that the queue is ``vertical'', with all of the vehicles queueing at a single point, queue length $\eta_i(\tilde{\at})$ does not correspond directly to the actual number of vehicles waiting in a queue at any specific time, but the overall effects of the queueing is correctly captured, and so is the total time loss.
The inflow to the queue $q^{\mathrm{in}}(\tilde{\at})$ is given as a delayed traffic flow through $\ax^{\mathrm{o}}$,
\begin{equation}
\label{eq:qin}
    q^{\mathrm{in}}(\tilde{\at}) = q_{\mathrm{o}}\left(t - \frac{\ax^{\w} - \ax^{\mathrm{o}}}{V^{\mathrm{ff}}}\right),
\end{equation}

\begin{equation}
\label{eq:qin}
    q^{\mathrm{in}}(\tilde{\at}) =
    q^{\mathrm{queue}}(\at)=
    q_{\mathrm{o}}\left(t - \frac{\ax^{ego}(t) - \ax^{\mathrm{o}}}{V^{\mathrm{ff}}}\right),
\end{equation}

and the outflow from the queue $q^{\mathrm{out}}(\tilde{\at})$ is given as
\begin{equation}
    q^{\mathrm{out}}(\tilde{\at}) = \begin{cases}
    \min\{q^{\mathrm{in}}(\tilde{\at}), q^{\mathrm{cap}}(\tilde{\at})\},& \eta(\tilde{\at}) = 0, \\
    q^{\mathrm{cap}}(\tilde{\at}),& \eta(\tilde{\at}) > 0,
    \end{cases}
\end{equation}
where the queue capacity $q^{\mathrm{cap}}(\tilde{\at})$ depends on the stage of the breakdown,
\begin{equation}
    q^{\mathrm{cap}}(\tilde{\at}) = \begin{cases}
    Q^{\max},& \tilde{\at} < \tilde{t}^{\mathrm{b}},\\
    Q^{\mathrm{b}},& \tilde{t}^{\mathrm{b}}\le \tilde{\at} < \tilde{t}^{\mathrm{t}},\\
    Q^{\mathrm{t}},& \tilde{t}^{\mathrm{t}} \le \tilde{\at} < t^{\w}, \\
    Q^{\max}, & \tilde{\at} \ge t^{\w}.
    \end{cases}
\end{equation}
Here, $Q^{\mathrm{b}}$ is the bottleneck capacity when the ego vehicle is stationary (either in the shoulder lane, or in one of the traffic lanes), and $Q^{\mathrm{t}}$ is the moving bottleneck capacity while the ego vehicle is being towed.
If $\eta({\at}^{\w})>0$ the queue remains even after the towed ego vehicle leaves the road at $t=t^{\w}$, until all the remaining congestion is dissipated at road capacity rate $Q^{\max}$.
For this purpose, we let the position of the ego vehicle nominally be $x_{\mathrm{ego}}(t) = x^{\w}$ for $t \ge t^{\w}$.
Note that since $x_{\mathrm{ego}}(t^{\w}) = x^{\w}$, we have $\tilde{t}^{\w} = t^{\w}$.

Finally, substituting \eqref{eq:qout} and \eqref{eq:qin} into \eqref{eq:timeloss_def}, it is straightforward to show that the total public time loss is given by
\begin{equation}
    \ptl = \int_{\tilde{t}^{\mathrm{b}}}^{t^{\mathrm{d}}} \eta(\tilde{\at}) \mathrm{d}\tilde{\at},
\end{equation}
 given the evolution of the queue length $\eta(\tilde{\at})$ from $\tilde{t}^{\mathrm{b}}$ to the time when the congestion is dissipated $t^{\mathrm{d}}$.

This expressions can be significantly simplified if we assume that the traffic inflow to the link $e$ is approximately constant $q_{\mathrm{o}}(t) = q_{\mathrm{o}} =\mathrm{const}$, and that $q_{\mathrm{o}}>Q^{\mathrm{b}}$, $q_{\mathrm{o}}>Q^{\mathrm{t}}$.
In this case, the queue length is piecewise-linear in time,
\begin{equation}
    \eta(\tilde{\at}) = \begin{cases}
    0,
    & \tilde{\at} \le \tilde{\at}^{\mathrm{b}},\\
    (q_e-Q^{\mathrm{b}})(\tilde{\at} - \tilde{\at}^{\mathrm{b}}), 
    & \tilde{\at}^{\mathrm{b}}< \tilde{\at} \le \tilde{\at}^{\mathrm{t}}, \\
    \eta(\tilde{\at}^{\mathrm{t}}) + (q_e - Q^{\mathrm{t}})(\tilde{\at}-\tilde{\at}^{\mathrm{t}}),
    & \tilde{\at}^{\mathrm{t}} < \tilde{\at} \le \at^{\w},\\
    \eta(\at^{\w}) - (Q^{\max} - q_e)(\tilde{\at} - \at^{\w}), 
    & \at^{\w} < \tilde{\at} \le \at^{\mathrm{d}}, \\
    0, & \tilde{\at}> \at^{\mathrm{d}},
    \end{cases}
\end{equation}
the congestion dissipates at time
\begin{equation}
    t^{\mathrm{d}} = \frac{Q^{\max} t^{\w} - q_e \tilde{t}^{\mathrm{b}} - Q^{\mathrm{b}} (\tilde{t}^{\mathrm{t}} - \tilde{t}^{\mathrm{b}}) - Q^{\mathrm{t}} (t^{\w} - \tilde{t}^{\mathrm{t}})}{Q_e^{\max} - q_e}
\end{equation}
and the total time loss is given by
\begin{equation}
\begin{split}
\ptl = & q_e \frac{
{t^{\mathrm{d}}}^2 - {\tilde{t}^{\mathrm{b}}}{\vphantom{\tilde{t}^a}}^2 }{2} - Q^{\mathrm{b}} \frac{
{\tilde{t}^{\mathrm{t}}}{\vphantom{\tilde{t}^a}}^2 - {\tilde{t}^{\mathrm{b}}}{\vphantom{\tilde{t}^a}}^2
}{2} - \\
& Q^{\mathrm{t}} \frac{
{t^{\w}}^2 - {\tilde{t}^{\mathrm{t}}}{\vphantom{\tilde{t}^a}}^2
}{2} - 
Q^{\max} \frac{
{t^{\mathrm{d}}}^2 - {t^{\w}}^2
}{2}. 
\end{split}    
\end{equation}

\subsubsection{Mission delay model}\label{sec4.2.2}
Given decision $r_i$, the time of mission delay can be approximated as below. 

For $ \di \in (0, D_i^{\mathrm{1}}) $, the ego vehicle breaks down after driving a distance of $\di$, after which the worst-case strategy is applied, the time of mission delay $c_{2}(i,\di)$ is 
\begin{equation}
    c_{2}(i,\di) = \at^{\mathrm{w}}(i,\di) + \td^{\mathrm{m,b}}_{n^*} + \td^{\mathrm{wc}} _{n^*} - \at^{\mathrm{dl}} 
\end{equation}
\noindent where $\at^{\mathrm{w}}(i,\di)$ refers to the time the tow truck reaches the workshop. $\td^{\mathrm{m,b}}_{n^*}$ refers to the maintenance time of the broken down vehicle in workshop $n^*$, which is assumed to be provided by the workshop. $\td^{\mathrm{wc}}_{n^*}$ refers to the shortest time driving from workshop ${n^*}$ to the customer, which can be obtained by shortest path searching algorithms in $G$. $n^*$ and $\at^{\mathrm{w}}(i,\di)$ can be obtained by solving (1).   $\at^{\mathrm{dl}} $ refers to the delivery deadline.

For $ r_i^1 \in R^{\mathrm{w}}, \di \in (D_i^{\mathrm{1}}, +\infty) $, the ego vehicle reaches the workshop without breakdown. The time of mission delay is independent of $\di$ and can be computed as
\begin{equation}
    c_{2}(i,\di) =\at^{\mathrm{a}} + \td^{\mathrm{aw}}_{i} + \td^{\mathrm{m,f}}_{i} + \td^{\mathrm{wc}}_{i}-\at^{\mathrm{dl}}
\end{equation}
where $\td^{\mathrm{aw}}_{i}$ refers to the time driving from the alarm location to the planned workshop in $r_i$. $\td^{\mathrm{m,f}}_{i}$ refers to the maintenance time in the planned workshop in $r_i$ if the ego vehicle does not break down. $\td^{\mathrm{wc}}_{i}$ refers to the time driving from the planned workshop in $r_i$ to the customer. 

For For $ r_i^1 \in R^{\mathrm{cu}}, \di \in (D_i^{\mathrm{1}}, +\infty) $, the ego vehicle finishes the delivery without breakdown and there is no mission delay, i.e. $c_{2}(i,\di)=0$. 

\subsection{Decision assessment} \label{sec4.3}
\begin{figure}[ht]
\centering
\includegraphics[width=0.8\linewidth]{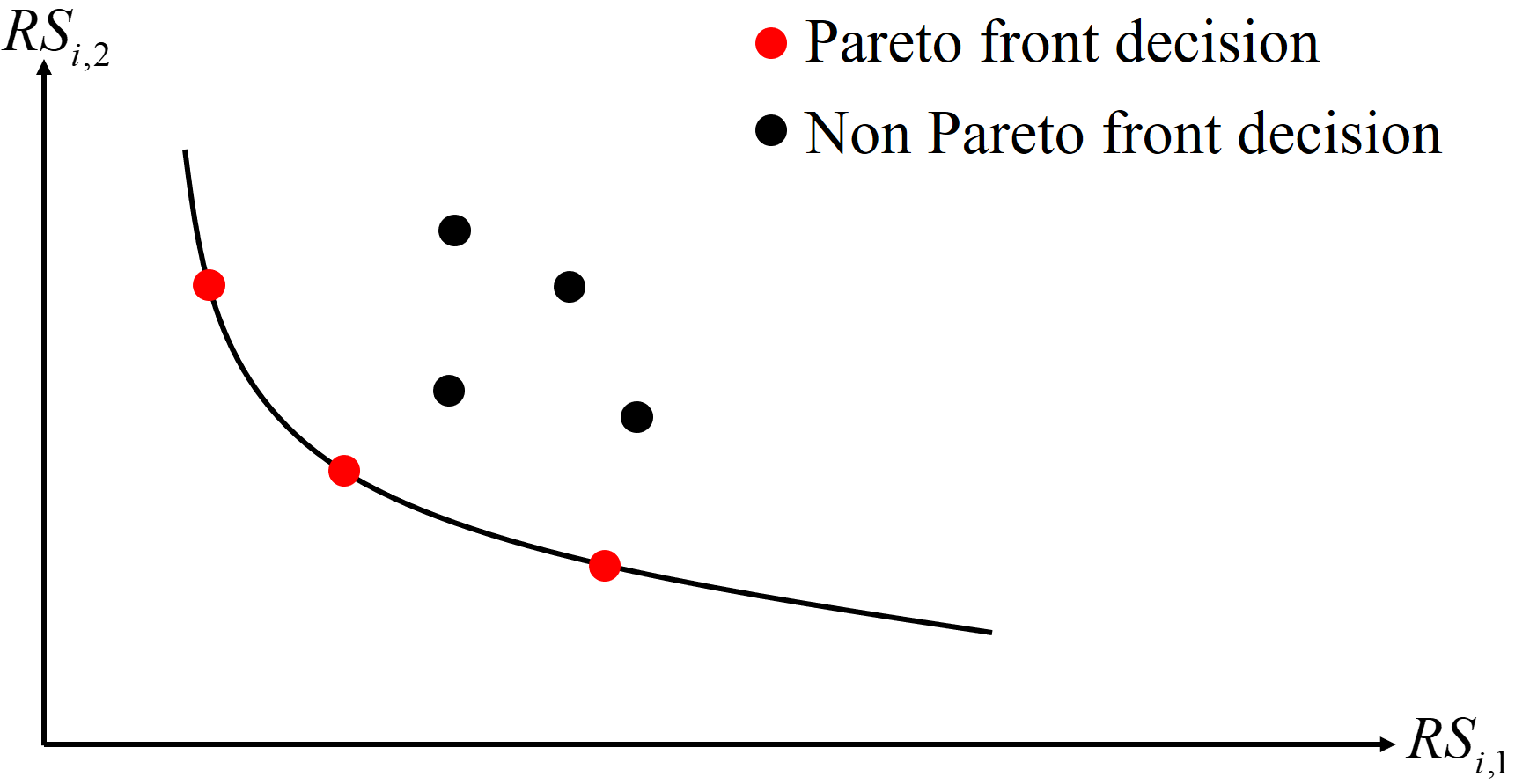}
\caption{Pareto front decisions.}
\label{fig:4}
\end{figure}

After risk assessment, we obtain the risk of public time loss $RS_1(i)$ and the risk of mission delay $RS_2(i)$ for each decision $r_i$. To further assess the decisions, we find the Pareto front, which contains all the non-dominated decisions that are superior to all the rest of the decisions in at least one criterion, shown in Fig. \ref{fig:4}. A decision $r_i$ strictly dominates another decisions $r_k$, denoted as $i \succ k$ if 
\begin{equation}
\begin{split}
    & RS_{1}(i)<RS_{1}(k) \hspace{1mm} \wedge \hspace{1mm} RS_{2}(i)<RS_{2}(k) \\ 
    &\textrm{for} \hspace{1mm} i\in [1,I], k\in[1,I], i \neq k.
\end{split}
\end{equation}

The Pareto front set, denoted as $PS(I)$ is
\begin{equation}
    PS(I) = \{i\in [1,I]: \{k\in [1,I]:k \succ i, k \neq i\} = \emptyset\}.
\end{equation}

\section{Numerical experiment}\label{sec5}

In this section, we demonstrate the MCDM problem and the proposed method with numerical experiments implemented in Matlab.

\subsection{Experiment setup} \label{sec5.1}
In this subsection, the experiment scenario is built and set as an intercity long-distance freight delivery. Note that the proposed method is generally applicable to urban transport scenarios and cars as well. The scenario built in this paper is for better relevance to the problem and for better heuristic interpretation since the vehicle runs mostly on motorways with much fewer route options.

\begin{figure*}[ht]
\centering
\includegraphics[width=\linewidth]{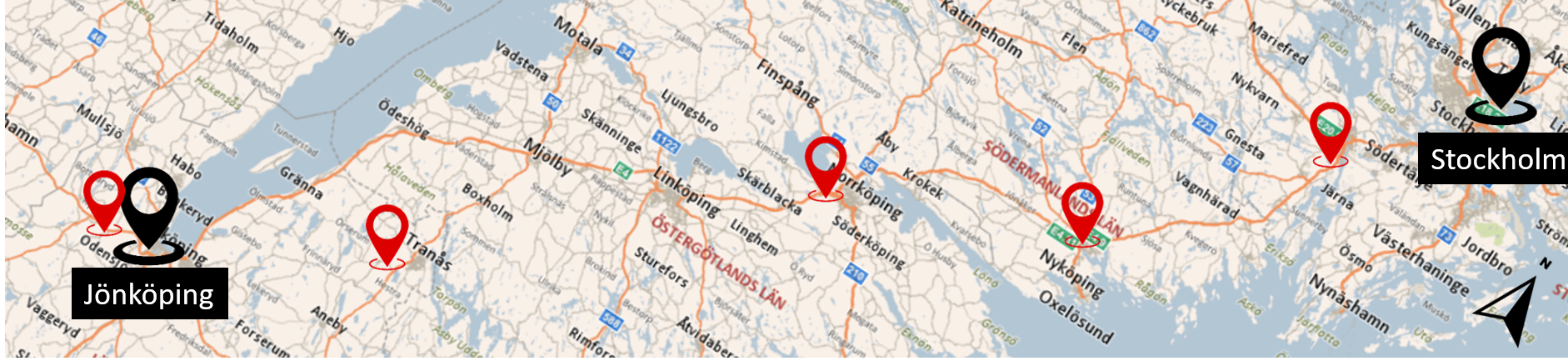}
\caption{The map of European motorway route E4 from Jönköping to Stockholm in Sweden. The red icons refer to available workshops that can conduct maintenance.}
\label{fig:5}
\end{figure*}

\begin{figure*}[ht]
\centering
\includegraphics[width=1\linewidth]{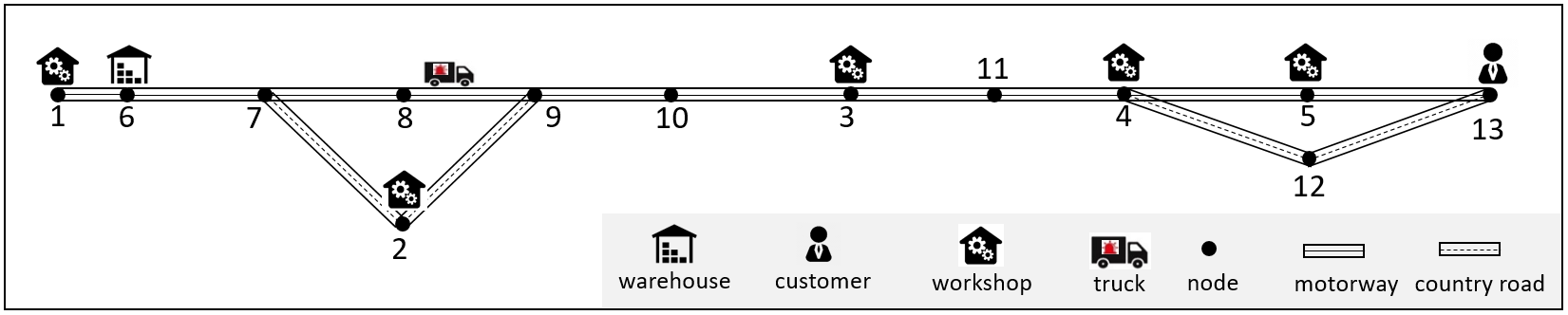}
\caption{Scenario of the intercity long-distance freight delivery mission.}
\label{fig:6}
\end{figure*}

The delivery scenario is derived from the Swedish freight transport practice. The map of the delivery scenario is shown in Fig. \ref{fig:5}. In this scenario, a truck, referred to as the ego truck, is running on European motorway route E4 between two major cities in Sweden, Jönköping and Stockholm. There are five workshops along the route affiliated to the truck company, marked with red icons. The scenario is simplified and sketched in Fig. \ref{fig:6}. The ego truck is driving from a warehouse in node 6 to the customer in node 13 with a total driving distance of 300 km. While driving on the planned motorway route, a fault is detected and the truck needs to make a decision. The five workshops are assigned with node index from 1 to 5. The workshop at node 2 is far from the motorway and the rest are close to the motorway. Despite the motorways between cities, there are also other types of roads connecting cities like country roads. Note that the real road network is more complex with more links and interchanges. We build the one in Fig. \ref{fig:6} with a balance of representativeness and simplicity for better heuristic interpretation. 

The configuration and properties of the road network are listed in TABLE \ref{tab:1}, including $e$ (link represented by node pairs), $L$ (link length), and  $RT(e)$ (road type). $RT(e)=1$ means that link $e$ is a motorway, and $RT(e)=0$ means that link $e$ is a country road. The free flow speed and towing speed on the motorways are set as 100 km/h and 30 km/h, respectively. The free flow speed and towing speed on country roads are set as 80 km/h and 30 km/h, respectively. 

{\renewcommand{\arraystretch}{1.3} 
\begin{table*}[ht]
 \centering
    \caption{Road Network configurations and properties}
    \begin{tabular}{|c| c| c| c| c| c| c| c| c| c| c| c| c| c| c|} \hline 
    $e$           &$(1,6)$ &$(2,7)$ &$(2,9)$ &$(3,10)$ &$(3,11)$ &$(4,5)$  &$(4,11)$ &$(4,12)$ &$(5,13)$ &$(6,7)$  &$(7,8)$ & $(8,9)$ & $(9,10)$ & $(12,13)$ \\
                  &$(6,1)$ &$(7,2)$ &$(9,2)$ &$(10,3)$ &$(11,3)$ &$(5,4)$  &$(11,4)$ &$(12,4)$ &$(13,5)$ &$(7,6)$  &$(8,7)$ & $(9,8)$ & $(10,9)$ & $(13,12)$ \\\hline
    $L$ (km)       &$15$    &$40$    &$40$     &$40$    &$30$     &$40$     &$30$     &$45$     &$40$     &$30$     &$30$    & $30$    & $30$     & $45$ \\\hline
    $RT(e)$       &$1$    &$0$    &$0$     &$1$    &$1$     &$1$     &$1$     &$0$     &$1$     &$1$     &$1$    & $1$    & $1$     & $0$ \\\hline
    \end{tabular}
    \label{tab:1}
\end{table*}
}

\begin{figure}
    \centering
    \includegraphics[width=1\linewidth]{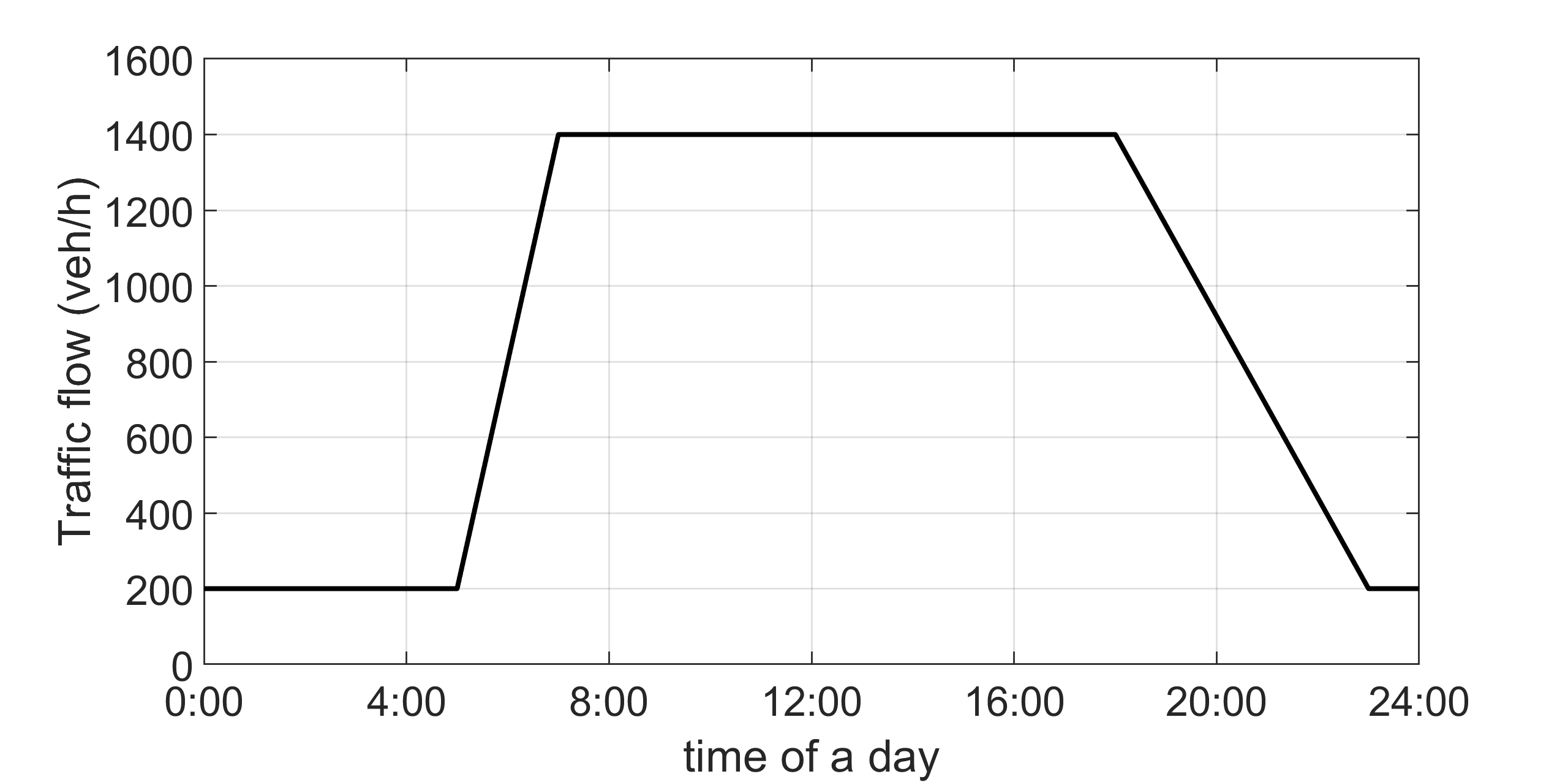}
    \caption{Traffic flow into the motorway link where the ego truck breaks down.}
    \label{fig:7}
\end{figure}

The traffic flow on the motorway in a day is shown in Fig. \ref{fig:7}, which is derived from the historical traffic flow statistics provided by the Swedish Transport Administration \cite{vagtrafikflodeskartan}. In practice, the traffic flow estimation is provided by traffic prediction models. Depending on the prediction method and capability, the traffic flow information may contain many more details, such as variation from day to day and variation from link to link. In this paper, for better heuristic interpretation, we simplify the traffic flow by only keeping the main trends and values, which highlight a sharp increase in traffic flow in the morning, i.e. the morning rush hour, and a gradual decrease in traffic flow in the evening. Such dynamic change in traffic flow in a day apply to most public roads. 

Based on this traffic flow model, we set the road capacity $Q^\textrm{max}=2000\hspace{1mm}$veh/h, the stationary bottleneck capacity $Q^\textrm{b}=800\hspace{1mm}$veh/h, and the moving bottleneck capacity $Q^\textrm{t}=1000\hspace{1mm}$veh/h. These parameters are road properties and can be obtained by simulation or field experiments. The setup in this paper mainly reflect that the road capacity is larger than the traffic inflow, so that there is no traffic congestion when the truck does not break down. The setup also corresponds to the situation that the ego truck breaks down where there is no shoulder, thus the capacity drop when a breakdown happens is significant.

\begin{figure}
    \centering
    \includegraphics[width=1\linewidth]{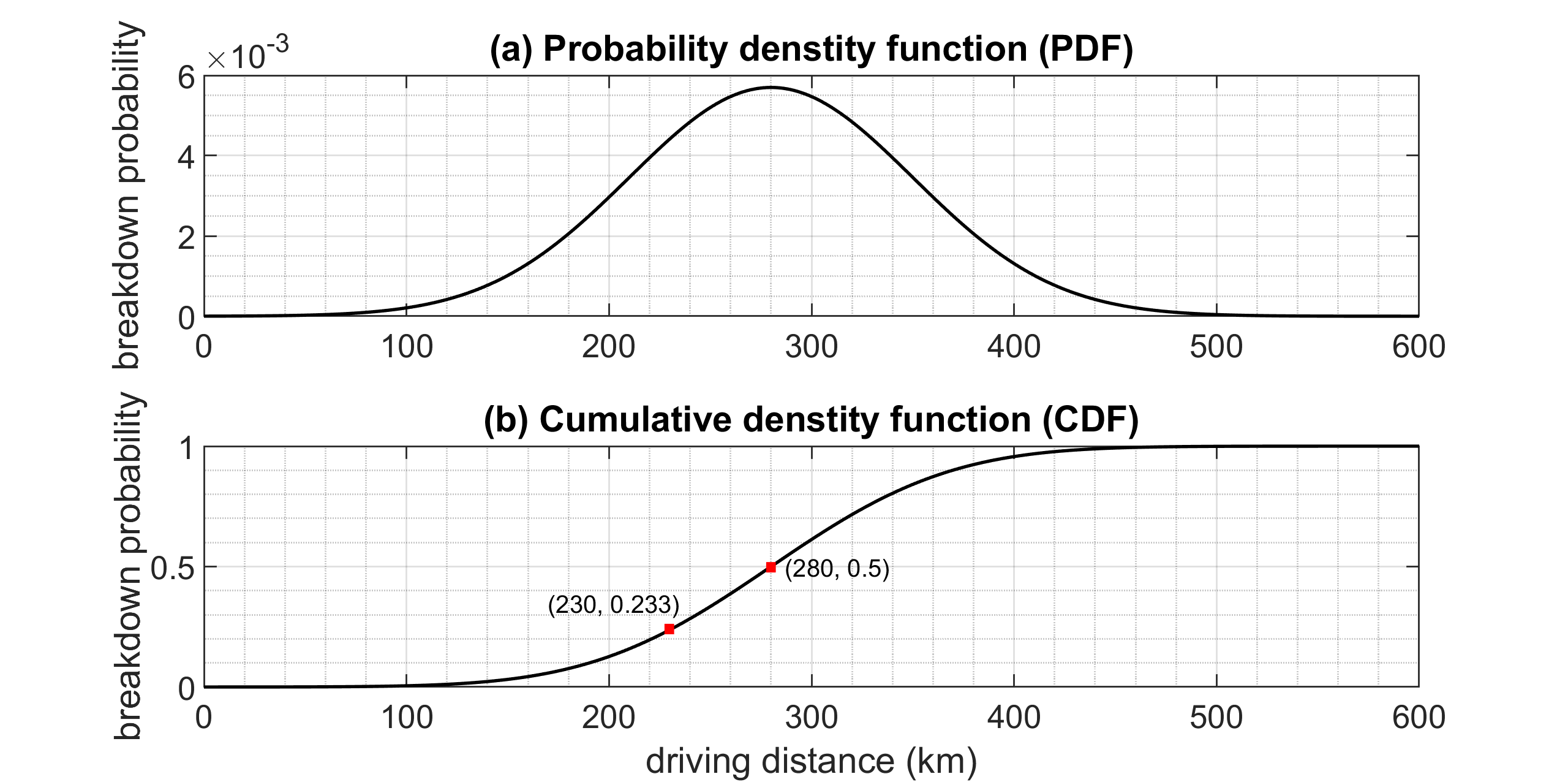}
    \caption{Probability distribution of the driving distance of the ego truck until breakdown.}
    \label{fig:8}
\end{figure}

We consider the scenario that a fault is detected when the ego truck is driving on link $(8,9)$ with 10 km to node 8, i.e. $n^{\textrm{ua}}=8, n^{\textrm{da}}=9$ and $d^\textrm{ua}= 10\hspace{1mm}$km. If the ego truck takes the planned route ($[9,10,3,11,4,5,13]$) to the customer, the driving distance to the customer is 230 km, taking 2.3 h at a free-flow speed of 100 km/h.

Assume that the driving distance of the ego truck to breakdown follows a normal distribution, i.e. $S \sim N(\mu, \sigma)$, where $\mu$ is the average driving distance and $\sigma$ is the variance. Note that the proposed decision-making model is generally applicable to any probability distribution of $S$ and normal distribution is used here as an example. To make the decision-making problem relevant, the probability of ego truck breakdown before finishing the delivery with the planned route should not be extremely high. Therefore, we set $\mu = 280$ and $\sigma = 70$. The probability density function and cumulative density function of $S$ are shown in Fig. \ref{fig:8}.

As shown in the figure, the ego truck is likely to break down after driving 100 km to 500 km, with a mean distance of 280 km. If it continues driving on the planned route with a distance of 230 km, it is likely to break down before reaching the customer with a probability of 0.233. With this setup, it is more likely to deliver without breakdown taking the planned route.

The maintenance time depends on various factors, such as the fault type and the workshop resources. We assume that this information is provided by the workshops in real-time. Here we set the maintenance time of a broken down truck $T_{n}^{m,b}=4\hspace{1mm}$h and that of a functioning truck $T_{n}^{m,f}=2\hspace{1mm}$h. This setup is mainly to reflect that it takes a much longer time to maintain a broken down vehicle, which makes a potential vehicle breakdown costly and unwanted. 

To this end, the scenario is set with all the necessary configurations. It is obvious that there are many parameters and many of them are provided by other vehicle modules or by other system actors in real time. In this paper, we set up these configurations mainly for the relevance of the problem, the heuristic interpretation, and the rationality of the correlation of different parameters. 

\subsection{Numerical experiment on the public time loss model} \label{sec5.2}

In this subsection, we demonstrate how the public time loss model evaluates the traffic congestion caused by a vehicle breakdown and the corresponding towing process by numerical experiment. Firstly we specify a determined breakdown scenario that: 
\begin{itemize}
    \item {the alarm time $T^{\mathrm{a}}$=10:00, indicating that the ego truck is and will be driving on motorways when the traffic flow is at their peak;}
    \item {the ego truck takes the planned route $[9,10,3,11,4,5,13]$ and breaks down after driving 110 km.}
\end{itemize}

\begin{figure}
    \centering
    \includegraphics[width=1\linewidth]{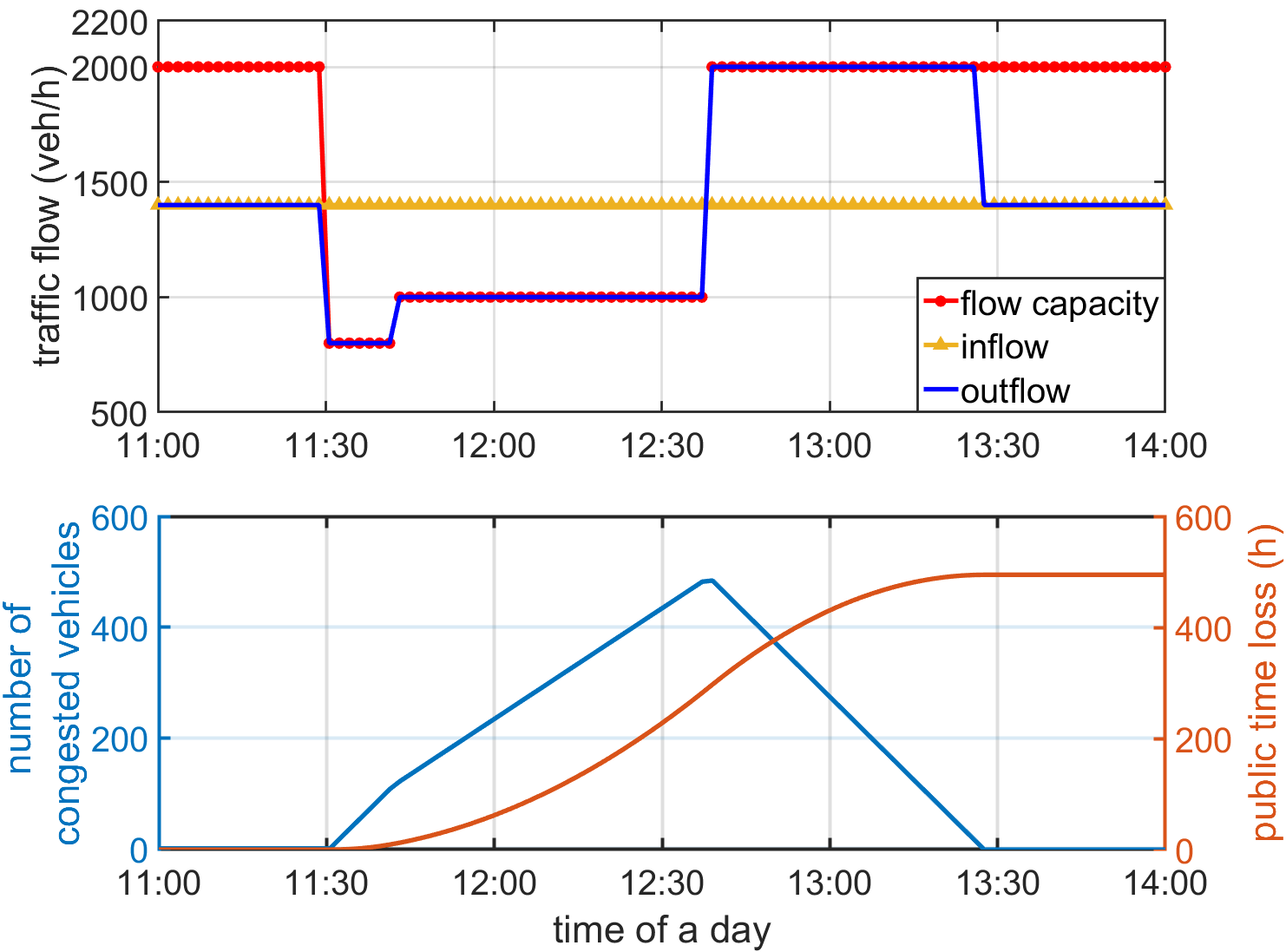}
    \caption{The traffic congestion caused by a vehicle breakdown and the corresponding towing process.} 
    \label{fig:9}
\end{figure}

With this setup, the ego truck breaks down on link $(3,11)$. Taking the worst-case strategy, the workshop at node 3 is selected to send out the tow truck and perform maintenance. With the public time loss model, we obtain the value of different times shown in TABLE \ref{tab:2} and the different values of traffic flow and traffic congestion shown in Fig. \ref{fig:9}.  

{\renewcommand{\arraystretch}{1.3} 
\begin{table}[!htbp]
 \centering
    \caption{The time of the occurrence of different events}
    \begin{tabular}{|c|c|c|} \hline
       Event         &  Absolute time & Relative time  \\ \hline
       Breakdown     &  $\at^\textrm{b}$=11:06  &  $\tilde{\at}^\textrm{b}$=11:30  \\ \hline
       Start towing  &  $\at^\textrm{t}$=11:18  &  $\tilde{\at}^\textrm{t}$=11:42  \\ \hline
       Exit motorway &  $\at^\textrm{e}$=12:38  &  $\tilde{\at}^\textrm{e}$=12:38  \\ \hline
    \end{tabular}
    \label{tab:2}
\end{table}
}

As shown in Fig. \ref{fig:9}, from 11:30 to 11:42, the ego truck breaks down and forms a stationary bottleneck, which causes a sharp decrease in the flow capacity. As a result, the number of congested vehicles increases quickly, as well as the public time loss. From 11:42 to 12:38, the tow truck tows the ego truck to the workshop at low speed and forms a moving bottleneck. During this time, the flow capacity increases while still being less than the inflow. As a result, the number of congested vehicles and the public time loss continue to increase at a lower rate. After the tow truck exits the motorway at 12:38, the flow capacity gains to the normal level and surpasses the inflow. During this time, the congestion is discharged quickly, shown as a sharp decrease in the number of congested vehicles. At 13:28, all the congested vehicles is discharged, and the traffic regains normal. 

From this case, we can see that although the capacity drop of a moving bottleneck is less than that of a stationary bottleneck, the impact on traffic congestion is significant since the low-speed towing process takes a long time. This also applies to the congestion discharging process after the tow truck exits the motorway.

Taking this specific case with a determined breakdown situation as an example, we demonstrate the applicability and the effectiveness of the public time loss model. 

\subsection{Numerical experiments on the risk evaluation model}\label{sec5.3}

In this subsection, we demonstrate the applicability and the effectiveness of the risk evaluation model with two cases set up as below.

In \textit{Case 1}, the time when the fault is detected is set as $T^{\mathrm{a}}$=10:00 and the mission deadline is set as $T^\textrm{dl}$=13:00, during which the traffic flow is at peak.

Using the decision-alternative generation method in Section \ref{sec4.1}, in total 9 decisions are generated, shown in TABLE \ref{tab:3}. Among them, the first 7 decisions are driving to a workshop first, and the latter 2 decisions are driving to the customer first. 

{\renewcommand{\arraystretch}{1.3} 
\begin{table}[!htbp]
 \centering
    \caption{decision alternatives }
    \begin{tabular}{|c|c|c|c|} \hline
       $i$ &  route of the first mission $r_i^{\textrm{1}}$ & route of the second mission $r_i^{\textrm{2}}$ \\ \hline
       $1$ &  [9,2,7,6,1]             & [1,6,7,8,9,10,3,11,4,5,13]   \\ \hline
       $2$ &  [9,8,7,6,1]             & [1,6,7,8,9,10,3,11,4,5,13] \\ \hline
       $3$ &  [9,2]                   & [2,9,10,3,11,4,5,13]\\ \hline
       $4$ &  [9,8,7,2]               & [2,9,10,3,11,4,5,13] \\ \hline
       $5$ &  [9,10,3]                & [3,11,4,5,13] \\ \hline
       $6$ &  [9,10,3,11,4]           & [4,5,13] \\ \hline
       $7$ &  [9,10,3,11,4,5]         & [5,13] \\ \hline
       $8$ &  [9,10,3,11,4,5,13]      & [13,12,4] \\ \hline
       $9$&   [9,10,3,11,4,12,13]     & [13,12,4]  \\ \hline
    \end{tabular}
    \label{tab:3}
\end{table}
}

In \textit{Case 2}, the scenario and the parameter configuration are the same as in \textit{Case 1}, except that the time when the fault is detected is set as $T^{\mathrm{a}}$=18:00 and the mission deadline is set as $T^\textrm{dl}$=21:00. Same as \textit{Case 1}, there are 3 h from the alarm time until the mission deadline. The decision alternatives of this case are the same as that of \textit{Case 1} shown in TABLE \ref{tab:3}.

With the setup of these two cases, the risk of public time loss $RS_{1}(i)$ and the risk of mission delay $RS_{2}(i)$ of the decisions of each case are listed in TABLE \ref{tab:4}. As shown in the table, in general, the risk of public time loss $RS_{1}(i)$ of \textit{Case 2} is much lower than that of \textit{Case 1}. Intuitively, this reflects the main difference between the setups of the two cases correctly. In \textit{Case 2}, for the first 200 km the ego truck may drive, the probability of breakdown is 0.12, which is considerably low. It takes 2 h on average to drive for 200 km on motorways. After 2 h from the alarm time, it is 20:00 and the traffic flow on motorways has decreasing sharply to 900 veh/h, which is close to the stationary bottleneck capacity and the moving bottleneck capacity. Therefore, in \textit{Case 2}, the ego truck will drive on public roads either with low breakdown probability or with low traffic flow, resulting in a low risk of public time loss. As for the risk of mission delay $RS_{2}(i)$, \textit{Case 1} and that of \textit{Case 2} have the same results. This is intuitively right since the difference in the setup of the two cases does not influence the delivery time. The comparison of \textit{Case 1} and \textit{Case 2} indicates that the public time loss caused by a vehicle breakdown depends heavily on the real-time traffic flow and the remaining useful life of the vehicle.

{\renewcommand{\arraystretch}{1.2} 
\begin{table}[!htbp]
 \centering
    \caption{Risk of public time loss $RS_{1}(i)$ and risk of mission delay $RS_{2}(i)$}
    \begin{tabular}{|c|c|c|c|c|} \hline
       decision index& \textit{Case 1}& \textit{Case 1}& \textit{Case 2}&\textit{Case 2} \\ \hline
       $i$ &  $RS_{1}(i)$ (h) & $RS_{2}(i)$ (h) &  $RS_{1}(i)$ (h) & $RS_{2}(i)$ (h)   \\ \hline
       $1$ &   9  & 3.9 & 2 & 3.9 \\ \hline
       $2$ &   6  & 3.4 & 3 & 3.4\\ \hline
       $3$ &   0  & 2.3 & 0 & 2.3\\ \hline
       $4$ &   2  & 2.9 & 1 & 2.9\\ \hline
       $5$ &   2  & 1.3 & 1 & 1.3\\ \hline
       $6$ &   12 & 1.4 & 4 & 1.4\\ \hline
       $7$ &   48 & 1.6 & 10 &1.6\\ \hline
       $8$ &   175& 1.2 & 14 &1.2\\ \hline
       $9$ &   12 & 1.7 & 4 & 1.7\\ \hline
    \end{tabular}
    \label{tab:4}
\end{table}
}

\begin{figure}[htp]
\includegraphics[clip,width=\columnwidth]{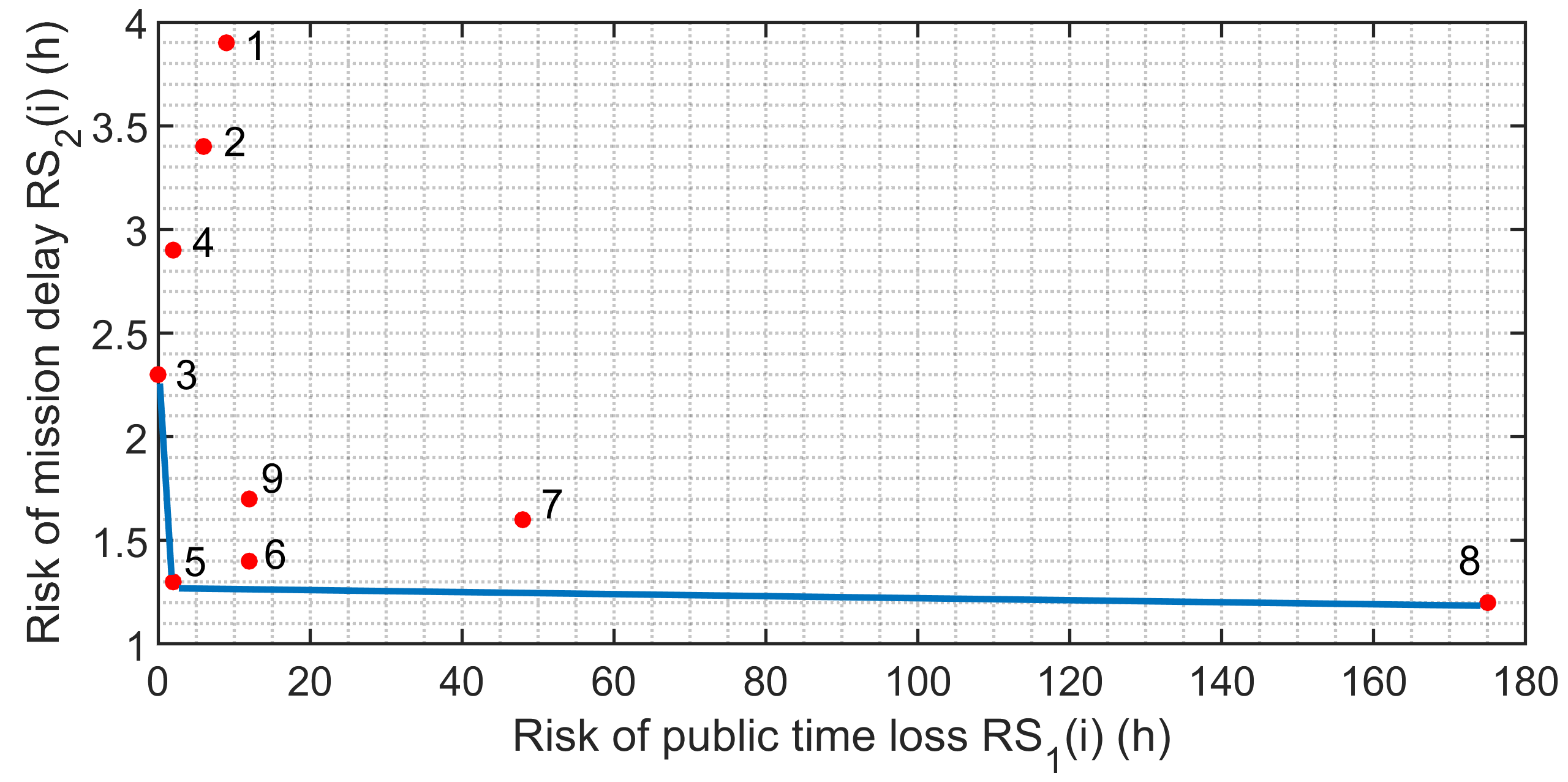}%
\caption{Pareto front decisions of \textit{Case 1}.}
\label{fig:10}
\end{figure}

The Pareto front decisions of \textit{Case 1} is shown in Fig. \ref{fig:10}. There are three elements in the Pareto front, decisions 3, 5, and 8, which are interpreted as below. 
\begin{itemize}
    \item {Decision 3, $\{[9,2],[2,9,10,3,11,4,5,13]\}$, has the lowest risk of public time loss of 0 h with a high risk of delivery delay of 2.3 h. This intuitively makes sense since the ego truck takes the fastest route to drive to the workshop for maintenance, thus having a low risk of breakdown and causing traffic congestion. However, it travels on links (9,2) and (2,9) and gets maintained in workshop at node 2, all of which take extra time before the delivery. Therefore, the risk of mission delay of this decision is relatively high. }
    \item {Decision 5, $\{[9,10,3],[3,11,4,5,13]\}$, has a low risk of public time loss of 2 h and a low risk of delivery delay of 1.3 h. This complies with intuition since the route to the workshop at node 3 is part of the route to the customer, thus causing less risk of mission delay than decision 3. Meanwhile, workshop at node 3 is the second closest to the ego truck among all the five workshops at the alarm time. Therefore, the ego truck has a very low risk of breakdown and causing traffic congestion.}     
    \item{Decision 8, $\{[9,10,3,11,4,5,13],[13,12,4]\}$, It has the highest risk of public time loss of 175 h, while the lowest risk of mission delay of 1.2 h. This is intuitively reasonable since the ego truck is likely to deliver without breakdown, thus having a low risk of delivery delay. However, once it breaks down, both the high traffic flow on the motorways and the long-distance driving would expose public transport to heavy traffic congestion.}
\end{itemize}

It is worth mentioning that in \textit{Case 1}, both decision 8 and decision 9 are delivery-first decisions. However, decision 9 has a risk of public time loss of 12 h, which is much less than that of decision 8. Different from decision 8, decision 9 avoids entering areas with dense traffic flows and takes country roads (links (4,12) and (12,13)) to the customer instead of motorways, where the traffic congestion caused by a breakdown is negligible. 

Taking today's practice, decision 8 is the optimal one since it minimizes the risk of mission delay for the private sector. Taking decision 8 as the baseline, for \textit{Case 1}, the change rate of the two types of risk by taking decisions 5, 3 and 9 are listed in TABLE \ref{tab:5}. In the table, $'+'$ refers to the growth rate, and $'-'$ refers to the reduction rate. 

The table shows that compared to decision 8, decisions 3, 5 and 9 all reduce the risk of public time loss to a large extent, with an increase of the risk of mission delay to different levels. Among them, decision 5 can be a good choice as a competitor to decision 8 since it only increases the risk of mission delay by 8$\%$. Although decision 9 is not an element in the Pareto front, if the private sector insists to deliver first, decision 9 can be recommended to reduce the risk of public time loss. 

{\renewcommand{\arraystretch}{1.2} 
\begin{table}[!htbp]
 \centering
    \caption{The change rate of the two types of risk in \textit{Case 1} }
    \begin{tabular}{|c|c|c|c|c|c|}\hline
       decision index& $i$                     &8&3&5&9\\\hline
       risk of public time loss &$RS_1(i)$ (h) & - & -100$\%$ & -99$\%$ & -93$\%$ \\\hline 
       risk of mission delay& $RS_2(i)$ (h)    & - & +92$\%$  & +8$\%$  & +42$\%$ \\\hline
    \end{tabular}\label{tab:shortest path}
    \label{tab:5}
\end{table}
}

With this numerical experiment, we illustrate that the proposed method can effectively obtain the Pareto front decisions. The Pareto front contains both maintenance-first and delivery-first decisions, which contributes to validating the relevance of the decision-making problem. The consistency between the experiment results and intuition contributes to validating the feasibility of the proposed method. The performance of the different decisions in the Pareto front contributes to validating the effectiveness of the proposed method. 

\subsection{Comparative study of two types of risk} \label{sec5.4}

In the previous subsection, the numerical experiment of \textit{Case 1} shows that in peak traffic, the risk of public time loss of a Pareto front decision can be very high. In this subsection, we conduct a comparative study of the risk of public time loss and the risk of mission delay by testing multiple cases. 

From the warehouse to the customer on the planned route, the fault might be detected at any location. On route [6,7,8,9,10,3,11,4,5] with a total distance of 260 km, we evenly take 52 samples of alarm locations at 5 km intervals. Each alarm location corresponds to a case indexed as $z \in [1,52]$. Alarm locations on link $(5,13)$ are not considered since the ego truck needs to drive to the customer at node 13 anyway and decision-making is not needed.

To make the decision-making problem relevant, two setups are adapted. First the alarm time is set as $T^\textrm{a}=10:00$ and the mission deadline is set as $T^\textrm{dl}= T^\textrm{a}+(300-5\cdot z)/100 + 0.5$. This setup specifies that if the ego truck drives to the customer following a motorway route without breakdown, it will reach the customer without delay. Secondly, the mean of the driving distance until the breakdown $s$ is set as $\mu = (300-5\cdot z)+50$, indicating that the ego truck can drive longer than the distance to the customer on average. This setup specifies that the probability of a vehicle breakdown is lower than 0.5 given $s$ follows a normal distribution. 

By simulation, we obtain the Pareto front of all the 52 cases. The results show that for all the cases there is more than one element in the Pareto front. This indicates that there is no unique dominating optimal decision if both risks are considered. Meanwhile, in the Pareto front of all the cases, there are both maintenance-first and delivery-first decisions. This indicates that both of these two distinctive types of decisions might perform well, at least in criterion. 

\begin{figure}
    \centering
    \includegraphics[width=1\linewidth]{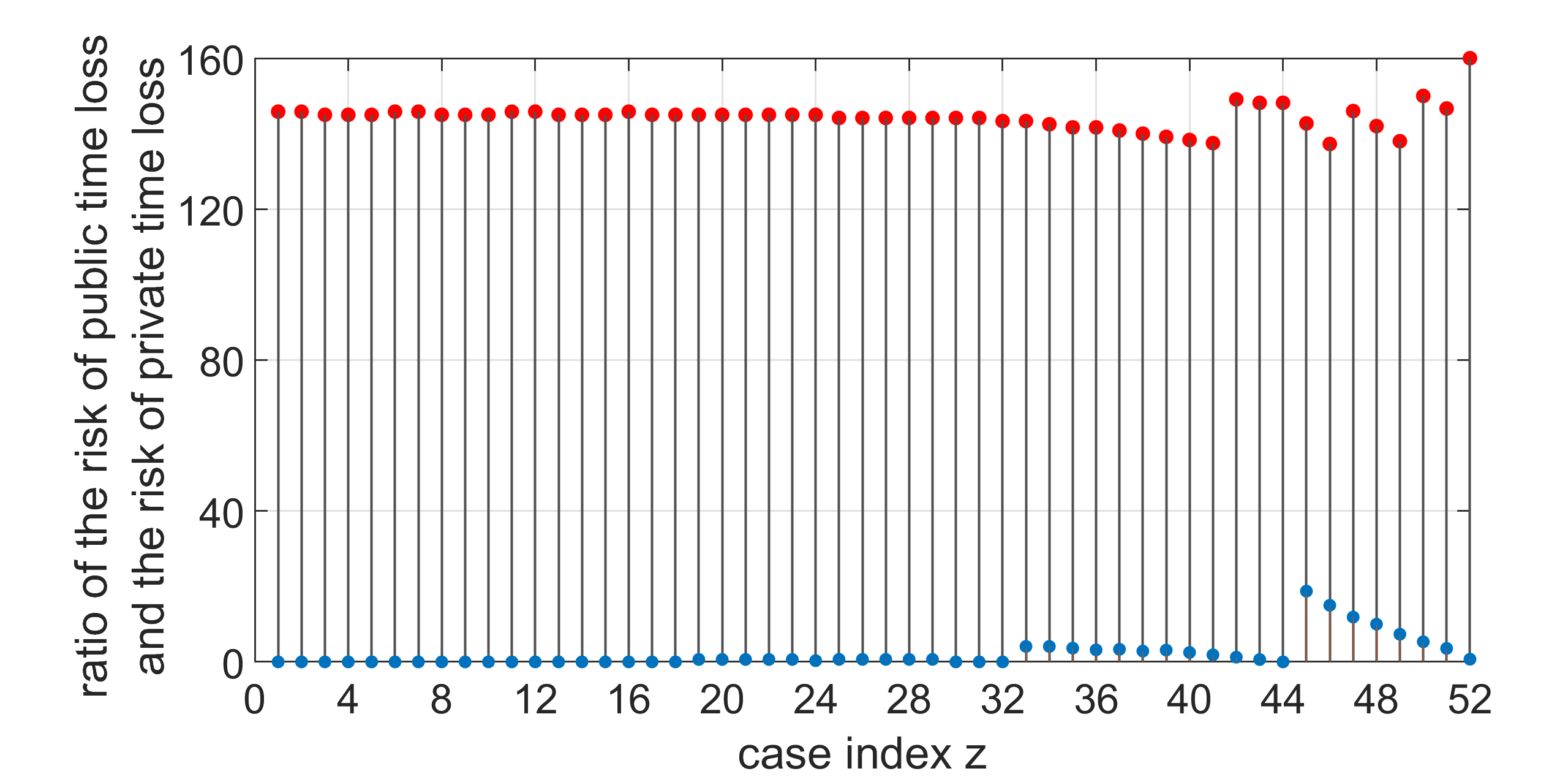}
    \caption{Ratio of the risk of public time loss and the risk of mission delay of each case. A red dot refers to the highest ratio of a case, while a blue dot refers to the lowest ratio of a case.}
    \label{fig:11}
\end{figure}

For each case, the ratio of the risk of public time loss and the risk of mission delay of the Pareto front decisions can be computed. The highest ratio and the lowest ratio of each case are shown in Fig. \ref{fig:11}. The highest ratios are around 150, while the lowest ratios of most cases are close to 0. Most of the highest ratios are from the delivery-first decisions. This indicates that although ignoring the risk of vehicle breakdown and continuing the delivery may reduce the risk of mission delay, it will cause a significant increase in the risk of public time loss in general.

Note that the ratios of some adjacent cases fluctuates suddenly, such as the lowest ratio of case 44 and case 45. In case 44, the ego truck is soon to arrive at node 4, where a workshops lies. In case 45, the ego truck has just passed node 4. Therefore, the lowest risk of public time loss in case 44 is close to zero, while that in case 45 is noticeable. 

\section{Discussion}\label{sec6}
\subsection{Application of the proposed method}\label{sec6.1}

In the previous section, the numerical experiments show that underestimating the risk of a vehicle breakdown on the road can expose the public transport to heavy traffic congestion and travel time loss. This could be an effective validation of the relevance of the problem considered in this paper. Using the proposed method, we can obtain the Pareto front decisions, as well as the other decisions that cause less public time loss. To this end, the main goals of this paper have been fulfilled. However, the way to further select the final decision that the vehicle executes is yet to be determined, which is essential for the proposed method to be applied. 

To solve this problem, we need to have a method to compare the public time loss and the private time loss or make trade-offs between them. In practice, this decision-making problem is handled by the private sector, where the societal loss is not considered. Furthermore, the two types of loss are subject to two distinctive stakeholders, namely the freight company and the public. As a result, there is no standard or reference method to make such a comparison or trade-off. Therefore, the best way to further select one decision is an open question. 

Challenging as it is, with increasing attention to collaborative decision-making within the public and private sectors in transportation, new business models and transport policies are likely to emerge, making the proposed method directly applicable in practice. 

% However, the economic loss caused by traffic congestion has attracted increasing attention \cite{nwokedi2020estimates}.

\subsection{Limitations}\label{sec6.2}
\subsubsection{Decision space}\label{sec6.2.1}
In real-world scenarios, the road network will be much more complex, and there will be many more feasible decisions, which may make the risk evaluation computationally expensive. To solve this problem, we can optimize the computation model or add additional constraints on the decision-alternative generation. In addition, this paper does not consider reducing driving speed after a fault is detected, which was considered in \cite{tao2022short}. However, the proposed method can be well adapted to consider reducing speed options in the same way as in \cite{tao2022short}. 

\subsubsection{Public time loss model}\label{sec6.2.2}
As a cross-disciplinary trial, the proposed public time loss model aims to build the connection between maintenance planning and traffic congestion management. It is preliminary and does not consider some complexities in traffic modeling, such as traffic flow overloading without vehicle disruption and traffic control after a vehicle breakdown. However, the proposed model is fundamental and can be well adapted and integrated into more complex traffic models. 

\subsubsection{Information acquisition}\label{sec6.2.3}
The proposed method requires various types of information as inputs, such as the fault information, the maintenance information, and the traffic information. Some information might be difficult to obtain in practice. For example, the remaining useful life of the vehicle under some fault conditions might be unavailable due to the absence of particular sensors and prognosis modules. Some information might be of low quality in practice, such as the real-time traffic flow prediction. Therefore, the applicability of the proposed method might be limited due to the absence or low quality of information input. However, the proposed method has build a comprehensive and fundamental basis for the proposed decision-making problem, with which further extension and elaboration can be applied.

\section{Conclusion and future work}\label{sec7}
In this paper, we formulate an MCDM problem for the maintenance planning of an operating vehicle under fault conditions considering both the public sector and the private sector. We propose a risk-based decision-making method to solve the problem and conduct various numerical experiments. The experiment results validate the relevance of the problem and the effectiveness of the method in reducing the risk of public time loss and the risk of mission delay. A comparative study of 52 cases validates the importance of considering the risk of public time loss in the decision-making of the private sector. These results contribute to increasing the awareness of public-private partnerships by collaborative decision-making when developing automated driving technologies. 

More efforts are needed to increase the generalization and applicability of the method in practice. In Section \ref{sec6.2}, the discussion on limitations indicates potential improvements in computation efficiency, the generality of the public time loss model, and the adaptation to more diversified information input. Furthermore, by adopting the proposed method in a transportation system of a larger scale and with a longer time horizon, the significance of considering the concerns of both the private sector and public sector in this decision-making process would be better validated.

\begin{comment}
\section^*{Acknowledgments}
This should be a simple paragraph before the References to thank those individuals and institutions who have supported your work on this article.
\end{comment}

\bibliographystyle{IEEEtran}
\bibliography{bib}{}

% Generated by IEEEtran.bst, version: 1.14 (2015/08/26)
\begin{thebibliography}{10}
\providecommand{\url}[1]{#1}
\csname url@samestyle\endcsname
\providecommand{\newblock}{\relax}
\providecommand{\bibinfo}[2]{#2}
\providecommand{\BIBentrySTDinterwordspacing}{\spaceskip=0pt\relax}
\providecommand{\BIBentryALTinterwordstretchfactor}{4}
\providecommand{\BIBentryALTinterwordspacing}{\spaceskip=\fontdimen2\font plus
\BIBentryALTinterwordstretchfactor\fontdimen3\font minus
  \fontdimen4\font\relax}
\providecommand{\BIBforeignlanguage}[2]{{%
\expandafter\ifx\csname l@#1\endcsname\relax
\typeout{** WARNING: IEEEtran.bst: No hyphenation pattern has been}%
\typeout{** loaded for the language `#1'. Using the pattern for}%
\typeout{** the default language instead.}%
\else
\language=\csname l@#1\endcsname
\fi
#2}}
\providecommand{\BIBdecl}{\relax}
\BIBdecl

\bibitem{realpe2015sensor}
M.~Realpe, B.~Vintimilla, and L.~Vlacic, ``Sensor fault detection and diagnosis
  for autonomous vehicles,'' in \emph{MATEC Web of Conferences}, vol.~30.\hskip
  1em plus 0.5em minus 0.4em\relax EDP Sciences, 2015, p. 04003.

\bibitem{ren2021traditional}
J.~Ren, M.~Green, and X.~Huang, ``From traditional to deep learning: Fault
  diagnosis for autonomous vehicles,'' in \emph{Learning Control}.\hskip 1em
  plus 0.5em minus 0.4em\relax Elsevier, 2021, pp. 205--219.

\bibitem{lu2013review}
L.~Lu, X.~Han, J.~Li, J.~Hua, and M.~Ouyang, ``A review on the key issues for
  lithium-ion battery management in electric vehicles,'' \emph{Journal of power
  sources}, vol. 226, pp. 272--288, 2013.

\bibitem{chen2019review}
H.~Chen and B.~Jiang, ``A review of fault detection and diagnosis for the
  traction system in high-speed trains,'' \emph{IEEE Transactions on
  Intelligent Transportation Systems}, vol.~21, no.~2, pp. 450--465, 2019.

\bibitem{shafi2018vehicle}
U.~Shafi, A.~Safi, A.~R. Shahid, S.~Ziauddin, and M.~Q. Saleem, ``Vehicle
  remote health monitoring and prognostic maintenance system,'' \emph{Journal
  of advanced transportation}, vol. 2018, 2018.

\bibitem{li2015prediction}
Z.~Li and Q.~He, ``Prediction of railcar remaining useful life by multiple data
  source fusion,'' \emph{IEEE Transactions on Intelligent Transportation
  Systems}, vol.~16, no.~4, pp. 2226--2235, 2015.

\bibitem{spooner1997fault}
J.~T. Spooner and K.~M. Passino, ``Fault-tolerant control for automated highway
  systems,'' \emph{IEEE Transactions on vehicular technology}, vol.~46, no.~3,
  pp. 770--785, 1997.

\bibitem{du2021fault}
S.~Du and S.~Razavi, ``Fault-tolerant control of variable speed limits for
  freeway work zone with recurrent sensor faults,'' \emph{IEEE Transactions on
  Intelligent Transportation Systems}, 2021.

\bibitem{TAOFAULT}
X.~Tao, J.~Lu, D.~Chen, and M.~Törngren, ``Probabilistic inference of fault
  condition of cyber-physical systems under uncertainty,'' \emph{IEEE Systems
  Journal}, vol.~14, no.~3, pp. 3256--3266, 2020.

\bibitem{de2020aging}
L.~De~Pascali, F.~Biral, and S.~Onori, ``Aging-aware optimal energy management
  control for a parallel hybrid vehicle based on electrochemical-degradation
  dynamics,'' \emph{IEEE Transactions on Vehicular Technology}, vol.~69,
  no.~10, pp. 10\,868--10\,878, 2020.

\bibitem{9613093}
L.~Rylander, M.~Eneberg, J.~Mårtensson, and A.~Pernestål, ``Design of
  diagnosis service system for self-driving vehicles - learnings from the
  driver’s role today,'' in \emph{2021 Global Reliability and Prognostics and
  Health Management (PHM-Nanjing)}, 2021, pp. 1--8.

\bibitem{TaxonomyAD}
``Taxonomy and definitions for terms related to driving automation systems for
  on-road motor vehicles.''

\bibitem{tao2022short}
X.~Tao, J.~M{\aa}rtensson, H.~Warnquist, and A.~Pernest{\aa}l, ``Short-term
  maintenance planning of autonomous trucks for minimizing economic risk,''
  \emph{Reliability Engineering \& System Safety}, vol. 220, p. 108251, 2022.

\bibitem{ma2017prioritizing}
X.~Ma, C.~Ding, S.~Luan, Y.~Wang, and Y.~Wang, ``Prioritizing influential
  factors for freeway incident clearance time prediction using the gradient
  boosting decision trees method,'' \emph{IEEE Transactions on Intelligent
  Transportation Systems}, vol.~18, no.~9, pp. 2303--2310, 2017.

\bibitem{chand2020analysis}
S.~Chand, E.~Moylan, S.~T. Waller, and V.~Dixit, ``Analysis of vehicle
  breakdown frequency: A case study of new south wales, australia,''
  \emph{Sustainability}, vol.~12, no.~19, p. 8244, 2020.

\bibitem{wang2005vehicle}
W.~Wang, H.~Chen, and M.~C. Bell, ``Vehicle breakdown duration modeling,''
  \emph{Journal of Transportation and Statistics}, vol.~8, no.~1, p.~75, 2005.

\bibitem{taha2018route}
A.-E. Taha and N.~AbuAli, ``Route planning considerations for autonomous
  vehicles,'' \emph{IEEE Communications Magazine}, vol.~56, no.~10, pp. 78--84,
  2018.

\bibitem{trid_2020}
\BIBentryALTinterwordspacing
``The future of autonomous vehicles: Global insights gained from multiple
  expert discussions,'' 2020-04. [Online]. Available:
  \url{https://trid.trb.org/view/1722481}
\BIBentrySTDinterwordspacing

\bibitem{wu2018optimizing}
J.~Wu, B.~Lin, H.~Wang, X.~Zhang, Z.~Wang, and J.~Wang, ``Optimizing the
  high-level maintenance planning problem of the electric multiple unit train
  using a modified particle swarm optimization algorithm,'' \emph{Symmetry},
  vol.~10, no.~8, p. 349, 2018.

\bibitem{ellefsen2019comprehensive}
A.~L. Ellefsen, V.~{\AE}s{\o}y, S.~Ushakov, and H.~Zhang, ``A comprehensive
  survey of prognostics and health management based on deep learning for
  autonomous ships,'' \emph{IEEE Transactions on Reliability}, vol.~68, no.~2,
  pp. 720--740, 2019.

\bibitem{nowakowski2018evolution}
T.~Nowakowski, A.~Tubis, and S.~Werbi{\'n}ska-Wojciechowska, ``Evolution of
  technical systems maintenance approaches--review and a case study,'' in
  \emph{International Conference on Intelligent Systems in Production
  Engineering and Maintenance}.\hskip 1em plus 0.5em minus 0.4em\relax
  Springer, 2018, pp. 161--174.

\bibitem{vskerlivc2020analysis}
S.~{\v{S}}kerli{\v{c}} and E.~Sokolovskij, ``Analysis of heavy truck
  maintenance issues,'' \emph{Pomorstvo}, vol.~34, no.~1, pp. 24--31, 2020.

\bibitem{tavares2016vehicles}
C.~M. Tavares and J.~Szpytko, ``Vehicles emerging technologies from maintenance
  perspective,'' \emph{IFAC-PapersOnLine}, vol.~49, no.~28, pp. 67--72, 2016.

\bibitem{voronov2020machine}
S.~Voronov, ``Machine learning models for predictive maintenance,'' Ph.D.
  dissertation, Link{\"o}ping University Electronic Press, 2020.

\bibitem{prytz2014machine}
R.~Prytz, ``Machine learning methods for vehicle predictive maintenance using
  off-board and on-board data,'' Ph.D. dissertation, Halmstad University Press,
  2014.

\bibitem{ahmadzadeh2014remaining}
F.~Ahmadzadeh and J.~Lundberg, ``Remaining useful life estimation,''
  \emph{International Journal of System Assurance Engineering and Management},
  vol.~5, no.~4, pp. 461--474, 2014.

\bibitem{bouvard2011condition}
K.~Bouvard, S.~Artus, C.~B{\'e}renguer, and V.~Cocquempot, ``Condition-based
  dynamic maintenance operations planning \& grouping. application to
  commercial heavy vehicles,'' \emph{Reliability Engineering \& System Safety},
  vol.~96, no.~6, pp. 601--610, 2011.

\bibitem{8790142}
Y.~{Wang}, S.~{Limmer}, M.~{Olhofer}, M.~T.~M. {Emmerich}, and T.~{Bäck},
  ``Vehicle fleet maintenance scheduling optimization by multi-objective
  evolutionary algorithms,'' in \emph{2019 IEEE Congress on Evolutionary
  Computation (CEC)}, 2019, pp. 442--449.

\bibitem{biteus2017planning}
J.~Biteus and T.~Lindgren, ``Planning flexible maintenance for heavy trucks
  using machine learning models, constraint programming, and route
  optimization,'' \emph{SAE International Journal of Materials and
  Manufacturing}, vol.~10, no.~3, pp. 306--315, 2017.

\bibitem{eglese2018disruption}
R.~Eglese and S.~Zambirinis, ``Disruption management in vehicle routing and
  scheduling for road freight transport: a review,'' \emph{Top}, vol.~26,
  no.~1, pp. 1--17, 2018.

\bibitem{ahmadi2013location}
A.~Ahmadi-Javid and A.~H. Seddighi, ``A location-routing problem with
  disruption risk,'' \emph{Transportation Research Part E: Logistics and
  Transportation Review}, vol.~53, pp. 63--82, 2013.

\bibitem{abu2021development}
A.~M. Abu-Monshar, A.~F. Al-Bazi, and Q.~H. Alsalami, ``On the development of a
  multi-layered agent-based heurisitc system for vehicle routing problem under
  random vehicle breakdown,'' \emph{Cihan University-Erbil Scientific Journal},
  vol.~5, no.~1, pp. 1--10, 2021.

\bibitem{dhahri2013vehicle}
A.~Dhahri, K.~Zidi, and K.~Ghedira, ``Vehicle routing problem with time windows
  under availability constraints,'' in \emph{2013 International Conference on
  Advanced Logistics and Transport}.\hskip 1em plus 0.5em minus 0.4em\relax
  IEEE, 2013, pp. 308--314.

\bibitem{li2020fifty}
Z.-C. Li, H.-J. Huang, and H.~Yang, ``Fifty years of the bottleneck model: A
  bibliometric review and future research directions,'' \emph{Transportation
  research part B: methodological}, vol. 139, pp. 311--342, 2020.

\bibitem{akhtar2021review}
M.~Akhtar and S.~Moridpour, ``A review of traffic congestion prediction using
  artificial intelligence,'' \emph{Journal of Advanced Transportation}, vol.
  2021, 2021.

\bibitem{nguyen2020traffic}
D.~Q. Nguyen-Phuoc, W.~Young, G.~Currie, and C.~De~Gruyter, ``Traffic
  congestion relief associated with public transport: state-of-the-art,''
  \emph{Public Transport}, vol.~12, no.~2, pp. 455--481, 2020.

\bibitem{khajeh2019traffic}
M.~Khajeh~Hosseini and A.~Talebpour, ``Traffic prediction using time-space
  diagram: A convolutional neural network approach,'' \emph{Transportation
  Research Record}, vol. 2673, no.~7, pp. 425--435, 2019.

\bibitem{chand2021examining}
S.~Chand, Z.~Li, V.~V. Dixit, and S.~T. Waller, ``Examining the macro-level
  factors affecting vehicle breakdown duration,'' \emph{International Journal
  of Transportation Science and Technology}, 2021.

\bibitem{bouziyane2020multiobjective}
B.~Bouziyane, B.~Dkhissi, and M.~Cherkaoui, ``Multiobjective optimization in
  delivering pharmaceutical products with disrupted vehicle routing problem,''
  \emph{International journal of industrial engineering computations}, vol.~11,
  no.~2, pp. 299--316, 2020.

\bibitem{aven2013uncertainty}
T.~Aven, P.~Baraldi, R.~Flage, and E.~Zio, \emph{Uncertainty in risk
  assessment: the representation and treatment of uncertainties by
  probabilistic and non-probabilistic methods}.\hskip 1em plus 0.5em minus
  0.4em\relax John Wiley \& Sons, 2013.

\bibitem{biswas2019multiobjective}
S.~Biswas, S.~G. Anavatti, and M.~A. Garratt, ``Multiobjective mission route
  planning problem: a neural network-based forecasting model for mission
  planning,'' \emph{IEEE Transactions on Intelligent Transportation Systems},
  vol.~22, no.~1, pp. 430--442, 2019.

\bibitem{tarjan1972depth}
R.~Tarjan, ``Depth-first search and linear graph algorithms,'' \emph{SIAM
  journal on computing}, vol.~1, no.~2, pp. 146--160, 1972.

\bibitem{rachmawati2020analysis}
D.~Rachmawati and L.~Gustin, ``Analysis of dijkstra’s algorithm and a*
  algorithm in shortest path problem,'' in \emph{Journal of Physics: Conference
  Series}, vol. 1566, no.~1.\hskip 1em plus 0.5em minus 0.4em\relax IOP
  Publishing, 2020, p. 012061.

\bibitem{cicic2021coordinating}
M.~{\v{C}}i{\v{c}}i{\'c}, X.~Xiong, L.~Jin, and K.~H. Johansson, ``Coordinating
  vehicle platoons for highway bottleneck decongestion and throughput
  improvement,'' \emph{IEEE Transactions on Intelligent Transportation
  Systems}, 2021.

\bibitem{vagtrafikflodeskartan}
\BIBentryALTinterwordspacing
``Vägtrafikflödeskartan,'' last accessed 2022-04-13. [Online]. Available:
  \url{https://vtf.trafikverket.se/SeTrafikinformation}
\BIBentrySTDinterwordspacing

\end{thebibliography}

\begin{IEEEbiography}[{\includegraphics[width=1in,height=1.25in,clip,keepaspectratio]{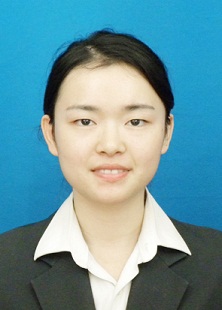}}]{Xin Tao}
received the M.Sc. degree in instrument science and technology from University of Science and Technology of China, Hefei, China, in 2017. She is currently pursuing the Ph.D. degree with the Integrated Transport Research Lab, KTH Royal Institute of Technology, Stockholm, Sweden. Her research activities focus on integrated vehicle health management, automated decision-making and uncertainty management.
\end{IEEEbiography}

\begin{IEEEbiography}[{\includegraphics[width=1in,height=1.25in,clip,keepaspectratio]{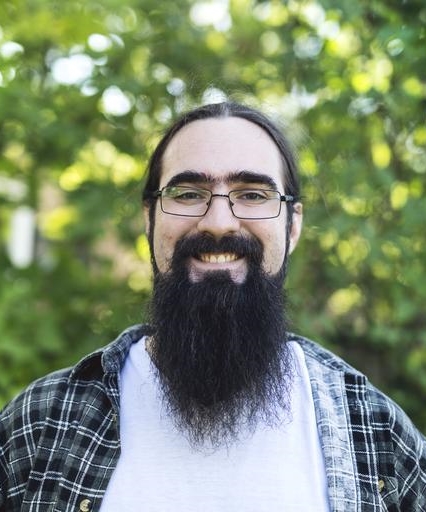}}]{Mladen \v{C}i\v{c}i\'{c}} received the M.Sc. degree in electrical engineering and computer science, from the School of Electrical Engineering, University of Belgrade, and the Ph.D. degree from KTH Royal Institute of Technology. He was a member of Marie Sk{\l}odowska Curie oCPS ITN, an affiliated Wallenberg AI, Autonomous Systems and Software Program (WASP) student, and a visiting scholar at the C2SMART University Transportation Center in the NYU Tandon School of Engineering. He is currently a post-doctoral researcher at GIPSA-lab, CNRS Grenoble. His research interests include modelling and control of mixed traffic.
\end{IEEEbiography}

\begin{IEEEbiography}[{\includegraphics[width=1in,height=1.25in,clip,keepaspectratio]{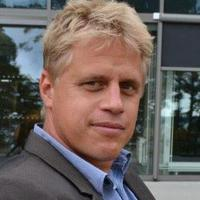}}]{Jonas M{\aa}rtensson}received the M.Sc. degree in vehicle engineering and the Ph.D. degree in automatic control from the KTH Royal Institute of Technology, Stockholm, Sweden, in 2002 and 2007, respectively. In 2016, he was appointed as a Docent. He is professor with the Division of Decision and Control Systems at KTH and he is the Director of the Integrated Transport Research Lab. He is affiliated faculty with the Wallenberg AI, Autonomous Systems and Software Program (WASP) and with Digital Futures.  His research interests are connected and automated transport systems, sustainable transport, heavy-duty vehicle platooning, path planning and predictive control of autonomous vehicles, support infrastructure and connectivity for automated vehicles, and related topics.
\end{IEEEbiography}

\vfill

\end{document}